# Antibiotic-dependent instability of homeostatic plasticity for growth and environmental load

*Shunnosuke Okada[1], Yudai Inabu[1], Hirokuni Miyamoto*[2,3,4,5], Kenta Suzuki[6], Tamotsu Kato[3], Atsushi Kurotani[7,8], Yutaka Taguchi[1], Ryoichi Fujino[1], Yuji Shiotsuka[1], Tetsuji Etoh[1], Naoko Tsuji[5], Makiko Matsuura[2,5], Arisa Tsuboi[4,5,7], Akira Saito[9], Hiroshi Masuya[6], Jun Kikuchi[7], Hiroshi Ohno[3]*, Hideyuki Takahashi[1]*.*

*Affiliations:*
[1]Kuju Agricultural Research Center, Graduate School of Agriculture, Kyushu University, Oita, Japan, 878-0201
[2]Graduate School of Horticulture, Chiba University, Chiba, Japan, 263-8522
[3]RIKEN Integrated Medical Science Center, Yokohama, Kanagawa, Japan, 230-004
[4]Japan Eco-science (Nikkan Kagaku) Co., Ltd., Chiba, Japan, 260-0034
[5]Sermas, Co., Ltd., Chiba, Japan, 271-8501
[6]RIKEN, BioResource Research Center, Tsukuba, Ibaraki, Japan, 305-0074
[7]RIKEN Center for Sustainable Resource Science, Yokohama, Kanagawa, Japan, 230-0045
[8]Research Center for Agricultural Information Technology, National Agriculture and Food Research Organization, Tsukuba, Ibaraki, Japan, 305-0856
[9]Feed-Livestock and Guidance Department, Dairy Technology Research Institute, The National Federation of Dairy Co-operative Associations (ZEN-RAKU-REN), Fukushima, Japan

* Cocorrespondence:
Hirokuni Miyamoto Ph.D., Chiba University, RIKEN IMS, Japan Eco-science Co., Ltd., and Sermas co. Ltd.,
E-mail: hirokuni.miyamoto@riken.jp, h-miyamoto@faculty.chiba-u.jp

Hiroshi Ohno Ph.D. and M.D., RIKEN IMS
E-mail: hiroshi.ohno@riken.jp

Hideyuki Takahashi, Kuju Agricultural Research Center, Graduate School of Agriculture, Kyushu University
Email: takahashi.hideyuki.990@m.kyushu-u.ac.jp

## Abstract
Reducing antibiotic usage in livestock animals has become an urgent issue worldwide to prevent antimicrobial resistance. Here, abuse of chlortetracycline (CTC), a versatile antibacterial agent, on the performance, blood components, fecal microbiota, and organic acid concentration in calves was investigated. Japanese Black calves were fed milk replacer containing CTC at 10 g/kg (CON) or 0 g/kg (EXP). Growth performance was not affected by CTC administration. However, CTC administration altered the correlation between fecal organic acids and bacterial genera. Machine learning methods such as association analysis, linear discriminant analysis, and energy landscape analysis revealed that CTC administration affected according to certain rules the population of various types of fecal bacteria. It is particularly interesting that the population of several methane-producing bacteria was high in the CON, and that of *Lachnospiraceae*, a butyrate-producing bacteria, was high in the EXP at 60 d of age. Furthermore, statistical causal inference based on machine learning data estimated that CTC treatment affects the entire intestinal environment, inhibiting butyrate production for growth and biological defense, which may be attributed to methanogens in feces. Thus, these observations highlight the multiple harmful impacts of antibiotics on intestinal health and the potential production of greenhouse gas in the calves.

*Running title*: Antibiotic dysbiosis against growth and environment

## Introduction
The beneficial effects of using antibiotics administered at non-therapeutic concentrations in feed as growth promoters (termed antimicrobial growth promoters, AGPs) were first recognized in the 1940s when chickens fed streptomycin exhibited enhanced growth and feed efficiency[1]. Since then, the growth-promoting effects of several antimicrobial agents have been documented in cattle, poultry, and swine[2]. In recent years, however, antimicrobial growth promoters (AGPs) have been under scrutiny from public health, food safety, and regulatory perspectives due to concerns about the potential for developing antimicrobial resistance (AMR)[3,4]. In addition, resistant bacteria against chlortetracycline (CTC), an antibacterial agent widely used in the livestock industry, have been discovered worldwide [5]. However, in the modern livestock industry, which depends on AGPs, removing AGPs from all feeds on the growth performance, nutrient metabolism, and intestinal environment in calves is relatively unknown.

The gut microbiota is known to have a role in shaping key aspects of postnatal life, such as the development of the immune system[6,7], and influencing the host's physiology, including energy balance. In addition, antibiotic usage has already been shown to affect fecal microbiota profiles in humans[8] and swine[9]. However, the effects of antibiotics on the gut microbiota, characterized by the use of a culture-independent metagenomic approach in ruminant species, particularly neonatal calves, have not been examined yet. Therefore, the objective of this study was to evaluate the effects of the lack of AGPs on the growth performance, health condition, and microbial diversity preweaning in beef calves.

Here, the physiological properties and bacterial population of the calves with and without antibiotics treatment were evaluated in detail. These data were evaluated by a procedure jointed with machine learning and statistical causal inference. As a result, statistical causal analyses based on machine learning data estimated the symbiotic bacterial groups that could potentially control the production of butyrate and methane. These observations provide critical viewpoints for developing sustainable livestock technologies to protect animal health and the global environment.

## Results and Discussion

To investigate how the lack of AGPs in the preweaning period affects the growth performance, health condition, and microbial diversity, calves were fed milk replacer containing CTC, an antibacterial agent widely used in the livestock industry, at 10 g/kg (CON) or 0 g/kg (EXP) (Fig. 1a). As a result, none of the calves developed major illnesses throughout the study. Administered feed to beef calves with or without AGPs did not adversely affect the phenotype of calves, as indicated by no difference in feed intake and body weight between treatments (Figs. 1b and S1ab). On the other hand, the contents of serum components, cytokines, hormones, fecal short-chain fatty acids, and phosphate did not appear to differ significantly between the two groups. (Figs. S2-S4). Differences in bacterial diversity and the major bacterial population were also not always observed between the treatment groups (Figs. S5-S7). However, the correlations between major fecal bacteria, organic acids, and phosphate were clearly different between the two groups (Figs. 1c and S8). In addition, linear discriminant analysis (LDA) effect size (LEfSe) analysis, a supervised machine learning method, revealed that CTC treatment affected the population of various types of fecal bacteria (Figs. 1d and S9). In addition, the correlation between the LDA-selected fecal bacterial population and organic acid and phosphate levels was altered by CTC treatment (Fig. S10), although these acid contents did not significantly differ.

Furthermore, association analysis, an unsupervised machine learning method, was performed for the items related to the presence or absence of CTC (Fig. S11). Although phenotypes were not affected by treatment, association analysis showed increased growth performance indicators, including BW, heart girth, waist circumference, and plasma IGF-1 concentration. These results did not contradict previous studies reporting the positive impact of antibiotics on growth in ruminants[10]. Statistical causal inference by LiNGAM (a Linear non-Gaussian Acylic Model) for each of the two groups estimated that the relationship of bacteria was different between the groups with and without CTC (Figs. S12 and S13). In particular, it suggested that CTC positively affected serum T-Cho concentration and negatively affected fecal butyrate and serum IgA production (Fig. S12a). Moreover, energy landscape analysis showed the entire stability of the physiological components in the experimental duration (Figs. 2a) but classified the unstable groups (four groups) after CTC exposure (Fig. S14). In addition, interaction network analyses estimated a negative effect of the genus *Methanobrevibacter* on the genus *Lactobacillus* abundance and fecal butyrate concentration (Fig. 2b). LiNGAM inferred a negative causality for butyrate production by antibiotic treatment and the genus *Methanobrevibacter* (Fig. 2c), together with a positive causality by genus *Dorea* belonging to family Lachnospiraceae.

As a result of association analysis, increased populations of the genera *Methanosphaera* and *Methanobrevibacter*, which belong to the family Methanobacteriaceae, were associated with the CTC treatment (Fig. S11), consistent with the results of LEfSe and energy landscape analysis. The increased population of methanogens is reported to be involved in developing obesity in humans[11]. In humans, methanogenic archaea were reported to be highly resistant to antibiotics[12], which may explain the increased methanogens in calves administered CTC. Enteric methane emissions are a problem due to energy loss, adverse effects on animal productivity, and environmental issues[11,13]. In addition, methane emissions from beef and dairy cows cause a loss of enteric methane energy, accounting for 2% to 12% of gross energy intake[14]. Thus, lipid accumulation due to an increased methanogen population can explain the relationship between antibiotic administration and growth.

Notably, we previously reported that *Caldibacillus hisashii*, a member of the class Bacilli, markedly reduces the abundance of the genus *Methanobrevibacter*[11]. Here, LiNGAM estimated that the class Bacilli works to diminish the effects of antibiotics (Fig. S15) and was causally related to the increase in fecal butyrate level and the abundance of the family Lachnospiraceae, a butyrate-producing bacterial family[15]. In addition, a rise in the fecal family Lachnospiraceae had positive causality with butyrate concentration. Butyrate has various beneficial effects, including improvement of intestinal development and barrier function, mitigation of inflammatory responses, and T cell-independent IgA response[16,17]. Based on these observations, it was inferred that the negative effect of antibiotic administration on fecal butyrate concentration was partly due to an increase in the abundance of methanogens, although the precise mechanism is unknown.

Thus, it is speculated that AGP administration has a negative impact on the gut function and performance of calves via a possible decrease in fecal butyrate production. These observations imply an instability of homeostatic plasticity for growth and environmental load by antibiotics. Overall, the current study first highlighted the possibility that the administration of antibiotics to promote the growth rate decreases intestinal butyrate production through alterations in the population of butyrate-producing bacteria and methanogens, which may have significant implications for the improvement of livestock productivity and reduction in the environmental load by antibiotics and global warming gas.

## Methods

***Animal management***

Twelve Japanese black calves (8 male and 4 female calves) were managed with the respective dams until 3 d of age. Thereafter, calves were separated from dams and moved to calf pens at 4 d of age. Milk replacer (Calf Top EX Black, Zen-Raku-Ren, Tokyo, Japan) was offered from 4 d of age using an automated calf feeder (Forster Technique, Co. Ltd., Germany). Calves were randomly assigned to one of two treatment groups [CON group: n = 6 (4 male and 2 female calves), initial BW = 31.5 ± 5.1 kg; EXP group: n = 6 (4 male and 2 female calves), initial body weight = 30.8 ± 5.1 kg]: calves in the CON group were fed milk replacer containing CTC at 10 g/kg, whereas those in the EXP group were fed antibiotic-free milk replacer. All milk replacers were formulated for 28.0% crude protein (CP), 18.0% crude fat (CF), and 108.0% total digestible nutrients (TDNs). The amount of milk replacer offered was increased from 0.5 kg/d to 1.0 kg/d from 4 to 21 d of age and maintained at this level until 60 d of age. All calves were fed the same calf starter containing 18.0% CP, 2.0% CF, and 72.5% TDNs (Hello Starter, Zen-Raku-Ren, Tokyo, Japan) and hay containing 12.4% CP, 4.1% CF, and 62.0% TDNs ad libitum from 4 d of age. All diets, except for the CON milk replacer, did not contain antibiotics. All calves had free access to water and mineral blocks (Cowstone A, Nippon Zenyaku kogyo Co. Ltd., Fukushima, Japan) throughout the study. Individual feed intake was recorded daily, and BW was measured every month. The present study was conducted at a Kuju agricultural research center (Oita, Japan), where Japanese black cattle were raised. All experimental protocols were approved by the Kyushu University Laboratory Animal Care and Use Committee (approval no. A30-355-1). The procedures used in the present study were performed according to ARRIVE guidelines and the Guidelines for Animal Experiments by the Faculty of Agriculture at Kyushu University.

***Sample collection***

On the day before sampling, access to the automated calf feeder was blocked. Blood samples were collected at 3, 30, and 60 d of age immediately before the morning feeding at 0900 h; this was achieved by using Vacutainers to collect serum (Venoject®II, Terumo Co. Ltd., Tokyo, Japan). All samples were centrifuged at 1200 ×g for 30 min at 20 °C, and serum was stored at -80 °C until analysis. Single-use gloves disinfected with 70% ethanol and sampling tubes with spoons were used to collect rectal feces from calves immediately after blood sampling at 3, 30, and 60 d of age. Fecal samples were stored at −30 °C and thereafter stored at −60 °C - −80 °C until analysis.

***Analyses of serum hormones and metabolites***

Serum concentrations of glucose, total cholesterol (T-Cho), nonesterified fatty acids (NEFAs), and β-hydroxybutyric acid (BHBA), urea nitrogen (SUN), and $Ca^{2+}$ were measured using commercially available assay kits by the glucose oxidase enzymatic method (glucose B-test; Wako Pure Chemical, Osaka, Japan), the acyl-CoA synthetase-acyl-CoA oxidase enzymatic method (FFAC; Wako Pure Chemical), ACS-ACOD method (NEFA C-test, Wako Pure Chemical Industries, Ltd., Japan), the bovine beta-hydroxybutyric acid (β-OHB) ELISA Kit (Huamei Biotech Co., Ltd., Wuhan, China), the enzymic method (DetectX Urea Nitrogen Colorimetric Detection Kit, Arbor Assays LLC, USA), and the OCPC method (Metallo Assays Calcium , Metallogenics Co, Ltd., Japan), respectively, according to the manufacturer's instructions. The serum concentration of insulin-like growth factor 1 (IGF-1) was measured by time-resolved fluoroimmunoassay as previously described[18,19].

**HPLC analysis for fecal metabolic acids**

The fecal samples (200-400 mg) were prepared according to a previous protocol[20] with some modifications. Briefly, the samples were mixed with a 9-fold volume of Milli-Q water for 10 min. After centrifugation at 15,000 rpm, all of the supernatants were filtered with 0.45 µm (Millex-HA Filter Unit SLHA025NB; Merck). The filtered solutions were subjected to high-performance liquid chromatography (HPLC) analysis. To determine the concentrations of lactic acid, acetic acid, propionic acid, butyric acid, valeric acid, isovaleric acid and phosphoric acid, frozen fresh fecal samples were analyzed by using an HPLC Prominence instrument (Organic Acid Analyzer; Shimadzu, Kyoto, Japan), which was equipped on an ion-exclusion column (Shim-pack SCR-102H; Shimadzu) and an electric conductivity detector (CDD-10AVP; Shimadzu). The analytical conditions were as follows: mobile phase, 5 mM *p*-toluenesulfonic acid; buffer, 5 mM *p*-toluenesulfonic acid, 20 mM Bis-Tris, and 0.2 mM EDTA-4H; temperature, 40 °C; and flow rate, 0.8 ml/min.

***Meta sequence analysis of bacterial 16S rRNA gene sequences***

The fecal samples from the CON (with antibiotic treatment) and EXP (without antibiotic treatment) groups were used for DNA extraction using a QIAamp PowerFecal DNA Kit (QIAGEN N.V. Inc.) according to the manufacturer's protocol. The DNA concentration was evaluated using a Quant-iTTM PicoGreen dsDNA Assay Kit (Thermo Fisher Scientific). The V4 region of the bacterial 16S rRNA gene (515F-806R) was sequenced according to a previous study[11,21]. The obtained sequences were filtered by Trimmomatic (http://www.usadellab.org/cms/?page=trimmomatic), and 5,000 trimmed reads per sample were analyzed with QIIME 1.9.1. The α-diversity, β-diversity, bacterial community, and correlations were visualized by using the packages "genefilter", "gplots", "ggplot2", "RColorBrewer", "pheatmap", "ape", "base", "dplyr", "easyGgplot2", "knitr", "ggthemes", "phyloseq", and "vegan" in R software (versions 4.0.5). The number of observed OTUs and Chao1, Shannon, and Simpson index values were assessed as measures of α-diversity.

The β-diversities were estimated by principal coordinate analysis (PCoA) using weighted or unweighted UniFrac distances based on the OTU distribution across samples. All 16S rRNA gene datasets were deposited in the GenBank Sequencing Read Archive database (accession number: DRA010973; BioSample numbers: SAMD00252058-SAMD00252093 / SSUB016265).

***LDA***

Linear discriminant analysis (LDA) is an elementary method of supervised machine learning. Here, LDA effect size (LEfSe) was used to identify genomic taxa characterizing the differences between the experimental conditions. LDA score plots and the cladogram based on LEfSe were visualized by Galaxy (https://huttenhower.sph.harvard.edu/galaxy/), which is described in a previous overview[22]. The populations of predominant bacterial community members (more than 0.1% of the total bacterial population) were analyzed. The threshold of the logarithmic LDA score for discriminative features was set at 3.0. A cladogram based on LEfSe and LDA score plots were made by the value for the factorial Kruskal–Wallis test among classes and the value for the pairwise Wilcoxon test between subclasses (set at 0.05) as nonparametric analyses. The pairwise comparisons among subclasses to be performed only among the subclasses with the same name were set at "Yes". The strategy for multiclass analysis was set as "All-against-all (more strict)".

**Energy landscape analysis**

The energy landscape analysis (ELA) was performed as previously described[23]. Energy landscape analysis is a data-driven method for constructing landscapes that explain the stability of community compositions across environmental gradients. Here, ELA is based on an extended pairwise maximum entropy model that explains the probability for the occurrence of the ecological state of sample $k$, $\sigma^{(k)}$ given the environmental condition $\epsilon^{(k)}$; as the ecological state, we combined the presence/absence status of selected taxa and levels of physiological factors, $\sigma^{(k)} = (\sigma_1^{(k)}, \sigma_2^{(k)}, \dots \sigma_N^{(k)}) = (\sigma_1^{(k)}, \dots, \sigma_{n_m}^{(k)}, \sigma_{N-n_c}^{(k)} \dots \sigma_N^{(k)})$ where $n_m$ is the number of bacterial taxa and $n_c$ is the number of chemicals; as the environmental condition, two environmental factors representing with (1) or without (0) antibiotic treatment ($\epsilon_{a,i}^{(k)}$) and the growth stages converted to the 0-1 range ($\epsilon_{s,i}^{(k)}$; 3, 30 and 60 days are converted as 0, 0.53 and 1, respectively) were combined as $\epsilon^{(k)} = (\epsilon_{a,i}^{(k)}, \epsilon_{s,i}^{(k)})$. The model can be written as:

(I) $$P(\sigma^{(k)}|\epsilon^{(k)}) = \frac{e^{-E(\sigma^{(k)}|\epsilon^{(k)})}}{\Sigma e^{-E(\sigma^{(k)}|\epsilon^{(k)})}},$$

(II) $$E(\sigma^{(k)}|\epsilon^{(k)}) = -(\Sigma_i \Sigma_j J_{ij} \sigma_i^{(k)} \sigma_j^{(k)} + \Sigma_i g_i^a \epsilon_{a,i}^{(k)} \sigma_i^{(k)} + \Sigma_i g_i^s \epsilon_{s,i}^{(k)} \sigma_i^{(k)} + \Sigma_i h_i \sigma_i^{(k)}).$$

Here, $P(\sigma^{(k)}|\epsilon)$ is the probability for the occurrence of an ecological state $\sigma^{(k)}$. Eq. (I) shows that the probability is high when energy $E(\sigma^{(k)}|\epsilon)$ is low and *vice versa*. In eq. (II), $E(\sigma^{(k)}|\epsilon)$ is defined as the sum of the effect of interaction among components, antibiotic treatment, growth stages, and the net effect of unobserved environmental factors. Parameters in eq. (II), namely, $J_{ij}$, $g_i^a$, $g_i^s$, and $h_i$, indicate the effect of the relationship among components ($J_{ij} > 0$ favors and $J_{ij} < 0$ disfavors the cooccurrence of components $i$ and $j$), the effect of the antibiotics on component $i$ (the antibiotic treatment positively ($g_i^a > 0$) or negatively ($g_i^a < 0$) affects the occurrence of component $i$), the effect of the growth stages on component $i$ (component $i$ favors the later ($g_i^s > 0$) or early ($g_i^s < 0$) growth stage) and how likely component $i$ occurs when the other factors are equal, respectively. We used a permutation test (N=2,000) [24] to obtain p-values for each parameter.

All the components in $\sigma^{(k)}$ were converted to the 0-1 range as follows. We first interpret the relative abundance of each bacterial genus in samples as presence (1) or absence (0) states by setting a threshold value of 0.001. Then, we selected 36 genera that appeared in more than 2 samples but fewer than 35 samples. We combined them into the density of eight selected chemicals, which were converted to the 0-1 range. Accordingly, for each of 36 samples, we obtained the set of explanatory variables $\sigma^{(k)}$ with 44 components, which accompanies the environmental condition $\epsilon^{(k)}$ representing the status for antibiotic treatment and growth stage.

### *DirectLiNGAM*

To estimate a structural model beyond the distribution of limited experimental data[21], the linear non-Gaussian acyclic model (LiNGAM) approach[25] involves independent component analysis and a non-Gaussian method for estimating causal structures. DirectLiNGAM was established with Python code on the website (https://github.com/cdt15/lingam) (Python version 3.6). The data calculated by the LiNGAM were visualized as LiNGAM networks in Gephi 0.9.2.

### *Statistical analyses*

In addition to LDA, association analysis, energy landscape analysis, and LiNGAM, individual data were analyzed using the MIXED procedure of JMP14 (SAS Institute Inc. Cary, NC, USA) according to the following model:

$$Y_{ijk} = \mu + G_i + T_j + C_k + GT_{ij} + e_{ijk}$$

where $Y_{ijk}$ is the dependent variable, $\mu$ is the overall mean, $G_i$ is the fixed effect of treatment, $T_j$ is the fixed effect of time (age) used as a repeated measure, $GT_{ij}$ is the fixed effect of treatment by time after birth interaction, $C_k$ is the random effect of the calf, and $e_{ijk}$ is the error term. A simple main effect test was performed to detect the differences between treatment groups at the same point. Significance was declared at $P < 0.05$, and a tendency was assumed at $0.05 \leq P < 0.20$. The relative values of dominant and/or characteristic bacteria were visualized through construction of a correlation graph and heatmap after the Pearson correlation coefficient was calculated for the selected bacteria (> 1% of the detected community and > 0.1% of the LDA selected detected community) by using R software (version 4.0.5). The data are presented as the means ± SDs.

## Acknowledgements

Special thanks to Ms. Chitose Ishii (Sermas co. ltd.) and Ms. Teruno Nakaguma (Sermas co. ltd.) for technical supports.

## Author Contribution

M.H., S.A., and T.H. conceived and designed the experiments; O.S., E.T., S.Y., F.R., and T.H. performed the farm experiments; O.S., K.T., T.N., and M.M. performed the biological and microbial assays; O.S., I.Y. M.H., K.T., S.K., K.A., T.N., M.M., T.A., and K.J. analyzed the raw data; O.S., I.Y. M.H., S.K., K.A., K.J., T.H. wrote the manuscript. All authors Reviewed the manuscript and approved the final manuscript.

## Competing interests

The authors declare no competing financial and non-financial interests.

## Data availability

The datasets presented in this study can be found as excel files (named DataS1 FINAL.xlsx) in the supplementary information. Furthermore, all 16S rRNA gene datasets are deposited in the DDBJ Sequence Read Archive and can be found as accession number: DRA010973. In addition, the R protocols for association analysis used in this study were deposited on the following websites (named Market Basket Analysis): http://dmar.riken.jp/Rscripts/ and http://dmar.riken.jp/NMRinformatics/.

## Figure and Table legends

**Fig. 1** (a) The experimental design. (b) Changes in BW during the period. (c) shows the correlation between fecal organic acids and bacterial genera during the period based on the sampling data at 30 d and 60 d. (d) Significant changes in the bacterial population calculated by LDA ($p<0.05$; > 3-fold change). CON: a group treated with antibiotics (n = 6); EXP: a group treated without antibiotics (n = 6).

**Fig. 2** (a) The entire energy landscape associated with antibiotic treatment are shown. The axis formed the energy landscape with compositional energy, community state, and treated time (days). (b) The interaction network shows the significant relationships in the extended pairwise maximum entropy model fitted to the observational data. The blue and red lines show positive and negative effects between the components, respectively. The components are selected by LDA and association analyses, respectively. The bacteria selected by both analyses were underlined. The abbreviations shows as follows: E_: family Erysipelotrichaceae; L_: family Lachnospiraceae; GLU: serum glucose; NEFA: serum nonesterified free fatty acid; Butyrate: fecal butyric acid; Propionate: fecal propionic acid; T-Cho: serum total cholesterol. (c) The calculated causal relationship of the components strongly linked with butyrate by DirectLiNGAM is visualized. The amounts of changes (Days 3, 30, and 60) with respect to values on Day 3 of the components lined with butyrate (green letters) were used for the calculation. The arrow shows a trend of the causal relationship. The number shows the value of the causal contribution calculated by DirectLiNGAM. The minus and plus value shows a negative and positive causal contribution, respectively.

## References

1 Moore, P. R., Evenson, A. & et al. Use of sulfasuxidine, streptothricin, and streptomycin in nutritional studies with the chick. *J Biol Chem* **165**, 437-441 (1946).

2 Cromwell, G. L. Why and how antibiotics are used in swine production. *Anim Biotechnol* **13**, 7-27, doi:10.1081/abio-120005767 (2002).

3 Woolhouse, M. E. & Ward, M. J. Microbiology. Sources of antimicrobial resistance. *Science* **341**, 1460-1461, doi:10.1126/science.1243444 (2013).

4 Van Boeckel, T. P. *et al.* Reducing antimicrobial use in food animals. *Science* **357**, 1350-1352, doi:10.1126/science.aao1495 (2017).

5 Barlow, G. M., Yu, A. & Mathur, R. Role of the Gut Microbiome in Obesity and Diabetes Mellitus. *Nutr Clin Pract* **30**, 787-797, doi:10.1177/0884533615609896 (2015).

6 Mazmanian, S. K., Liu, C. H., Tzianabos, A. O. & Kasper, D. L. An immunomodulatory molecule of symbiotic bacteria directs maturation of the host immune system. *Cell* **122**, 107-118, doi:10.1016/j.cell.2005.05.007 (2005).

7 Peterson, D. A., McNulty, N. P., Guruge, J. L. & Gordon, J. I. IgA response to symbiotic bacteria as a mediator of gut homeostasis. *Cell Host Microbe* **2**, 328-339, doi:10.1016/j.chom.2007.09.013 (2007).

8 Claesson, M. J. *et al.* Composition, variability, and temporal stability of the intestinal microbiota of the elderly. *Proc Natl Acad Sci U S A* **108 Suppl 1**, 4586-4591, doi:10.1073/pnas.1000097107 (2011).

9 Looft, T. *et al.* In-feed antibiotic effects on the swine intestinal microbiome. *Proc Natl Acad Sci U S A* **109**, 1691-1696, doi:10.1073/pnas.1120238109 (2012).

10 Bartley, E. E., Fountaine, F. C., Atkeson, F. W. & Fryer, H. C. Antibiotics in Dairy Cattle Nutrition. I. The Effect of an Aureomycin Product (Aurofac) on the Growth and Well-Being of Young Dairy Calves. *Journal of Dairy Science* **36**, 103-111, doi:10.3168/jds.s0022-0302(53)91466-0 (1953).

11 Inabu, Y. *et al.* Development of a novel feeding method for Japanese black calves with thermophile probiotics at postweaning. *J Appl Microbiol* **132**, 3870-3882, doi:10.1111/jam.15519 (2022).

12 Dridi, B., Fardeau, M. L., Ollivier, B., Raoult, D. & Drancourt, M. The antimicrobial resistance pattern of cultured human methanogens reflects the unique phylogenetic position of archaea. *J Antimicrob Chemother* **66**, 2038-2044, doi:10.1093/jac/dkr251 (2011).

13 Subepang, S., Suzuki, T., Phonbumrung, T. & Sommart, K. Enteric methane emissions, energy partitioning, and energetic efficiency of zebu beef cattle fed total mixed ration silage. *Asian-Australas J Anim Sci* **32**, 548-555, doi:10.5713/ajas.18.0433 (2019).

14 Gerber, P. J. *et al.* Technical options for the mitigation of direct methane and nitrous oxide emissions from livestock: a review. *Animal* **7 Suppl 2**, 220-234, doi:10.1017/s1751731113000876 (2013).

15 Zhang, J. *et al.* Beneficial effect of butyrate-producing Lachnospiraceae on stress-induced visceral hypersensitivity in rats. *J Gastroenterol Hepatol* **34**, 1368-1376, doi:10.1111/jgh.14536 (2019).

16 Isobe, J. *et al.* Commensal-bacteria-derived butyrate promotes the T-cell-independent IgA response in the colon. *Int Immunol* **32**, 243-258, doi:10.1093/intimm/dxz078 (2020).

17 Furusawa, Y. *et al.* Commensal microbe-derived butyrate induces the differentiation of colonic regulatory T cells. *Nature* **504**, 446-450, doi:10.1038/nature12721 (2013).

18 Sugino, T. *et al.* Effects of ghrelin on food intake and neuroendocrine function in sheep. *Anim Reprod Sci* **82-83**, 183-194, doi:10.1016/j.anireprosci.2004.05.001 (2004).

19 Laarman, A. H., Ruiz-Sanchez, A. L., Sugino, T., Guan, L. L. & Oba, M. Effects of feeding a calf starter on molecular adaptations in the ruminal epithelium and liver of Holstein dairy calves. *J Dairy Sci* **95**, 2585-2594, doi:10.3168/jds.2011-4788 (2012).

20 Poudel, P. *et al.* Development of a systematic feedback isolation approach for targeted strains from mixed culture systems. *J Biosci Bioeng* **123**, 63-70, doi:10.1016/j.jbiosc.2016.07.019 (2017).

21 Miyamoto, H. *et al.* A potential network structure of symbiotic bacteria involved in carbon and nitrogen metabolism of wood-utilizing insect larvae. *Sci Total Environ* **836**, 155520, doi:10.1016/j.scitotenv.2022.155520 (2022).

22 Segata, N. *et al.* Metagenomic biomarker discovery and explanation. *Genome Biol* **12**, R60, doi:10.1186/gb-2011-12-6-r60 (2011).

23 Suzuki, K., Nakaoka, S., Fukuda, S. & Masuya, H. Energy landscape analysis elucidates the multistability of ecological communities. *Ecological Monographs* **91**, e01469 (2021).

24 Harris, D. J. Inferring species interactions from co-occurrence data with Markov networks. *Ecology* **97**, 3308-3314 (2016).

25 Shimizu, S., Hoyer, P. O., Hyvarinen, A. & Kerminen, A. A Linear Non-Gaussian Acyclic Model for Causal Discovery. *Journal of Machine Learning Research*, 2003-2030 (2006).

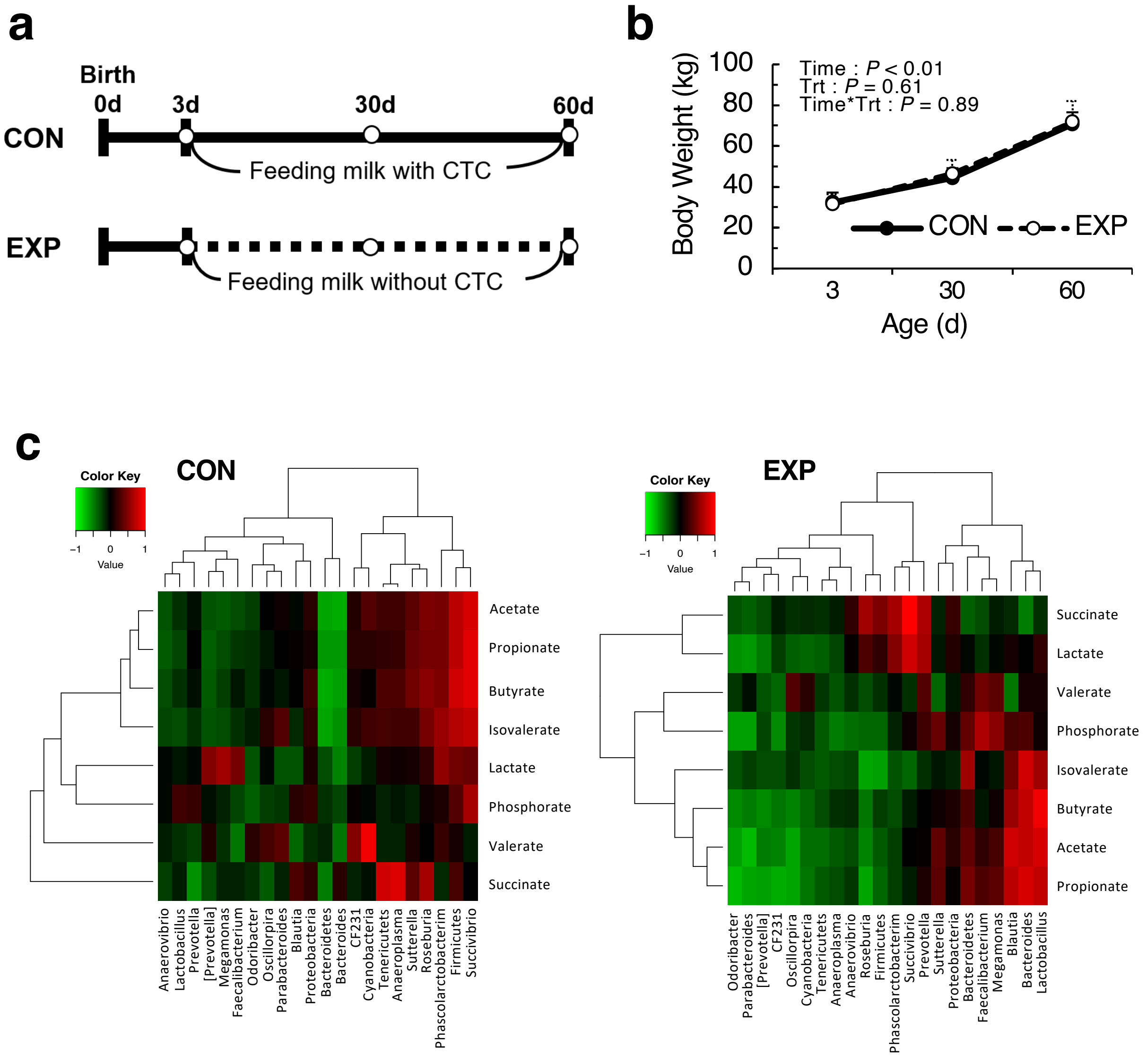

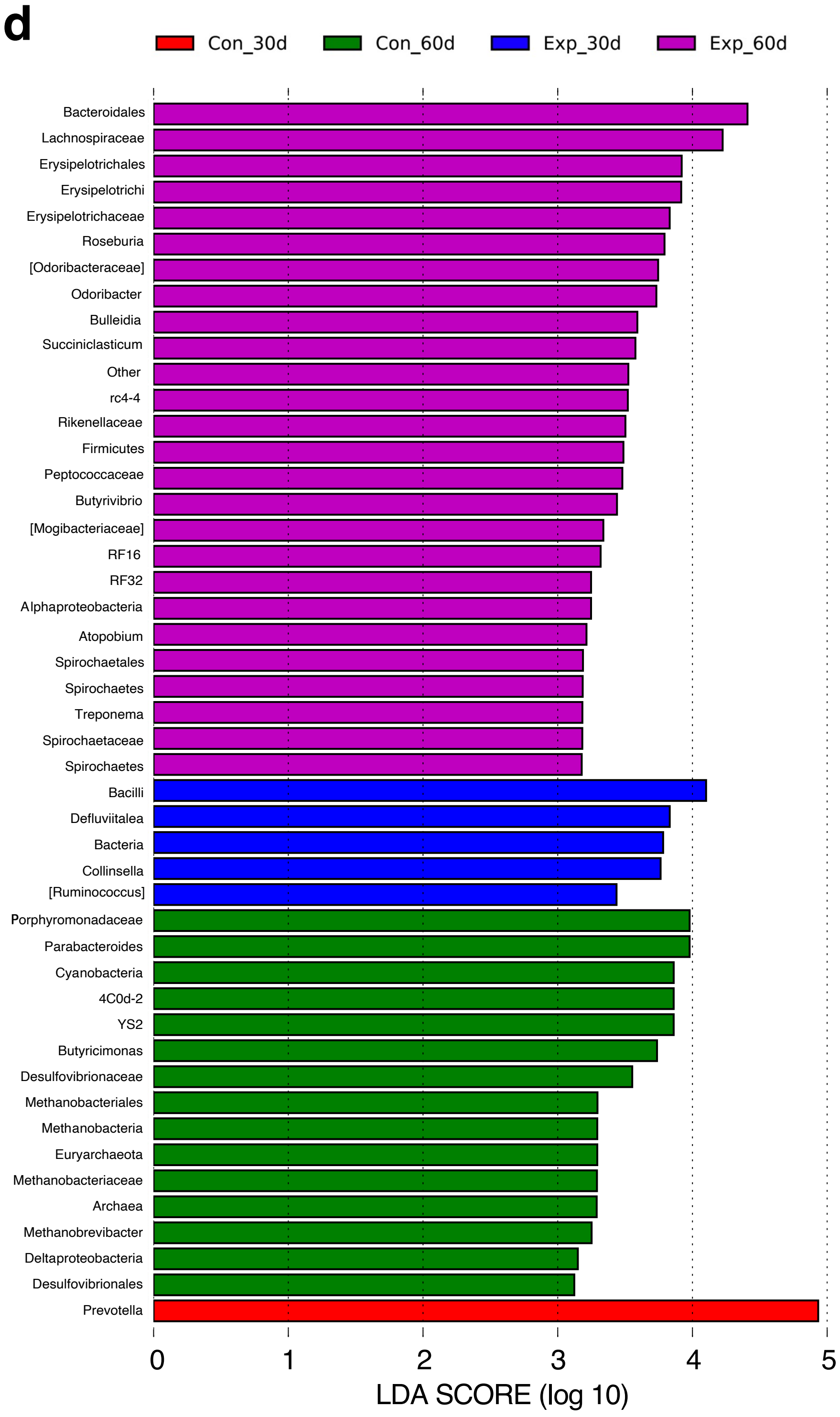

# Fig. 1

(a) The experimental design. (b) Changes in BW during the period. (c) shows the correlation between fecal organic acids and bacterial genera during the period based on the sampling data at 30 d and 60 d. (d) Significant changes in the bacterial population calculated by linear discriminant analysis (LDA) ($p<0.05$; > 3-fold change). CON: a group treated with antibiotics (n = 6); EXP: a group treated without antibiotics (n = 6).

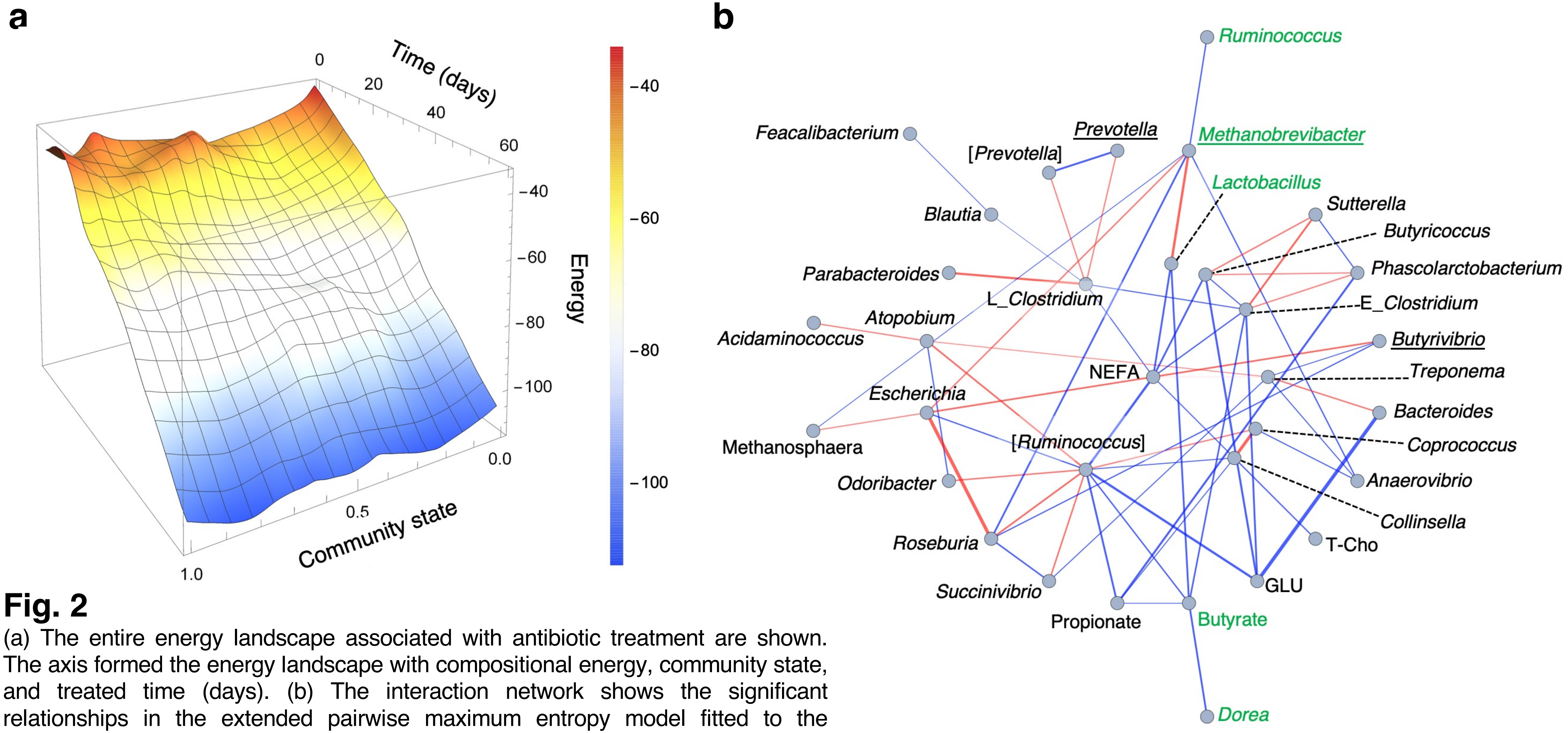

# Fig. 2

(a) The entire energy landscape associated with antibiotic treatment are shown. The axis formed the energy landscape with compositional energy, community state, and treated time (days). (b) The interaction network shows the significant relationships in the extended pairwise maximum entropy model fitted to the observational data.The blue and red lines show positive and negative effects between the components, respectively. The components are selected by LDA and association analyses, respectively. The bacteria selected by both analyses were underlined. The abbreviations shows as follows: E_: family Erysipelotrichaceae; L_: family Lachnospiraceae; GLU: serum glucose; NEFA: serum nonesterified free fatty acid; Butyrate: fecal butyric acid; Propionate: fecal propionic acid; T-Cho: serum total cholesterol. (c) The calculated causal relationship of the components strongly linked with butyrate by DirectLiNGAM is visualized. The amounts of changes (Days 3, 30, and 60) with respect to values on Day 3 of the components lined with butyrate (green letters) were used for the calculation. The arrow shows a trend of the causal relationship. The number shows the value of the causal contribution calculated by DirectLiNGAM. The minus and plus value shows a negative and positive causal contribution, respectively.

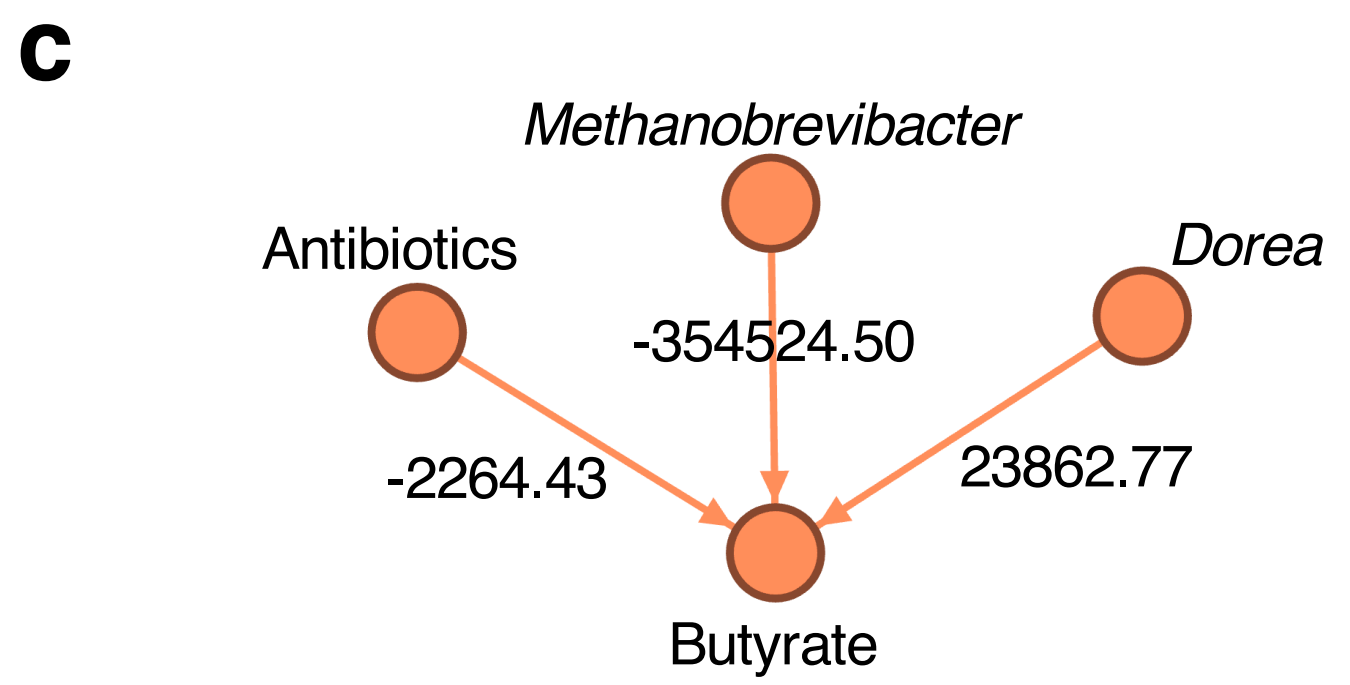

# Supplementary Information

## Antibiotic-dependent instability of homeostatic plasticity for growth and environmental load

*Shunnosuke Okada[1], Yudai Inabu[1], Hirokuni Miyamoto*[2,3,4,5], Kenta Suzuki[6], Tamotsu Kato[3], Atsushi Kurotani[7,8], Ryoichi Fujino[1], Yuji Shiotsuka[1], Tetsuji Etoh[1], Naoko Tsuji[5], Makiko Matsuura[2,5], Arisa Tsuboi[4,5,7], Akira Saito[9], Hiroshi Masuya[6], Jun Kikuchi[7], Hiroshi Ohno[3]*, Hideyuki Takahashi[1]*.*

[1]*Kuju Agricultural Research Center, Graduate School of Agriculture, Kyushu University, Oita, Japan, 878-0201*
[2]*Graduate School of Horticulture, Chiba University, Chiba, Japan, 263-8522*
[3]*RIKEN Integrated Medical Science Center, Yokohama, Kanagawa, Japan, 230-0045*
[4]*Japan Eco-science (Nikkan Kagaku) Co., Ltd., Chiba, Japan, 260-0034*
[5]*Sermas, Co., Ltd., Chiba, Japan, 271-8501*
[6]*RIKEN, BioResource Research Center, Tsukuba, Ibaraki, Japan, 305-0074*
[7]*RIKEN Center for Sustainable Resource Science, Yokohama, Kanagawa, Japan, 230-0045*
[8]Research Center for Agricultural Information Technology, National Agriculture and Food Research Organization, Tsukuba, Ibaraki, Japan, 305-0856
[9]*Feed-Livestock and Guidance Department, Dairy Technology Research Institute, The National Federation of Dairy Co-operative Associations (ZEN-RAKU-REN), Fukushima, Japan*

* Cocorrespondence:
Hirokuni Miyamoto Ph.D., hirokuni.miyamoto@riken.jp, h-miyamoto@faculty.chiba-u.jp
Hiroshi Ohno Ph.D. and M.D., RIKEN IMS, hiroshi.ohno @riken.jp
Hideyuki Takahashi Ph.D., takahashi.hideyuki.990@m.kyushu-u.ac.jp

**This file includes:** Figures. S1 to S15

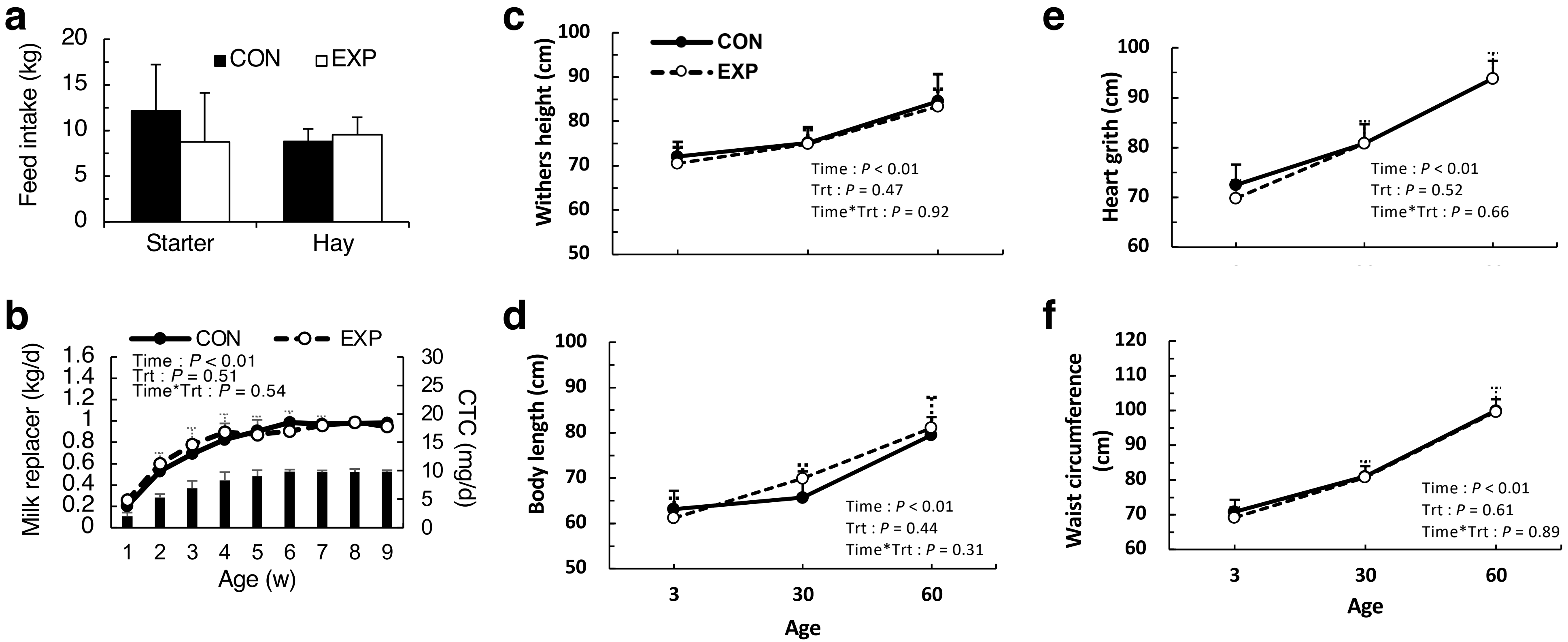

**Figure S1.** (a) feed intake and (b) antibiotic doses during the experimental period were shown, respectively. Doses of chlortetracycline (CTC) is expressed as bar graph. (d) shows changes in body weight during the period. Body height (c), body length (d), heart girth (e) and body waist circumference (f) at 3, 30, 60 days of age for the calves in fed milk replacer containing Chlortetracycline (CTC) at 10g/kg (CON) or 0g/kg (EXP). The values are means ± S.D.

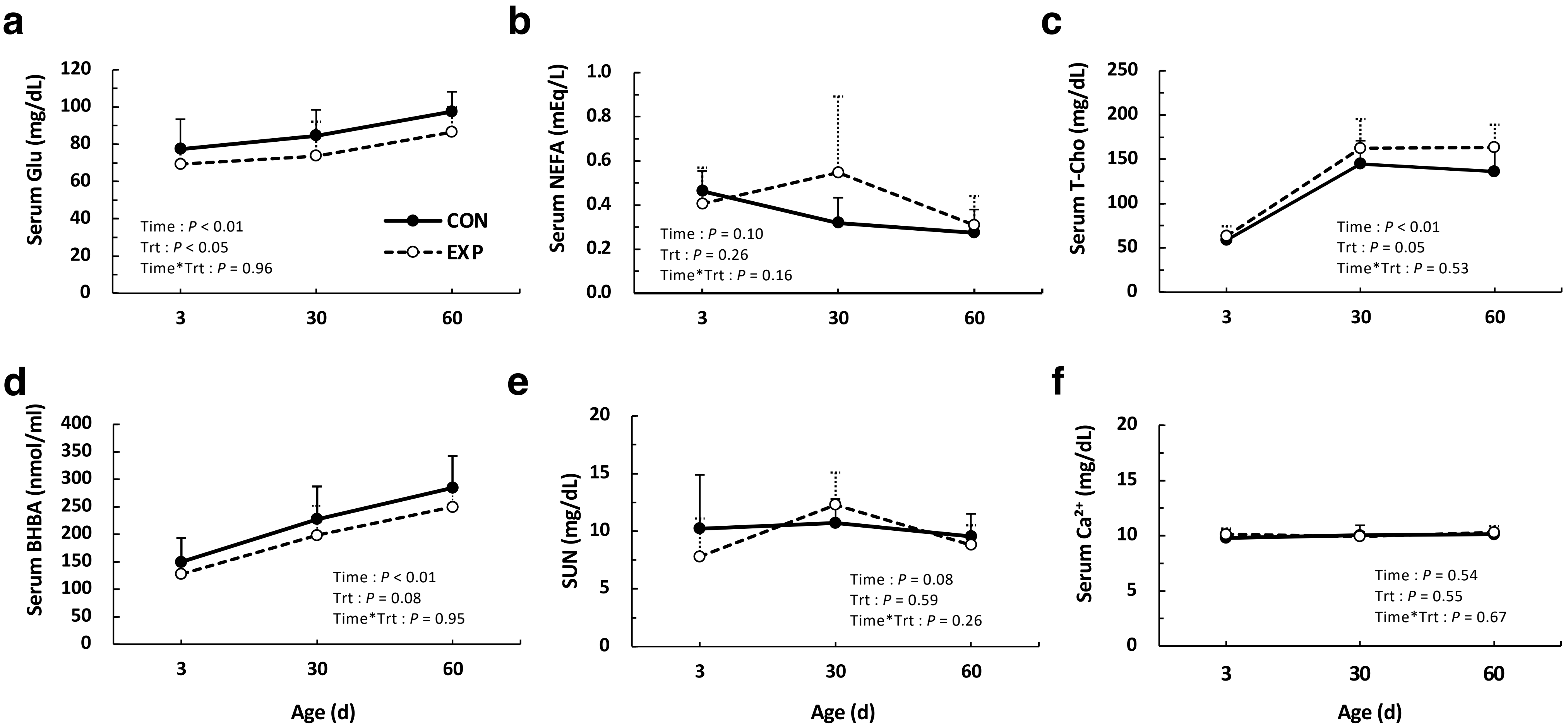

**Figure S2.** Blood components for calves fed milk replacer containing Chlortetracycline (CTC) at 10g/kg (CON) or 0g/kg (EXP). (a-f) Serum glucose (a), nonesterified fatty acid (NEFA) (b), total cholesterol (T-Cho) (c), β-hydroxybutyric acid (BHBA) (d), serum urea nitrogen (SUN) (e) and $Ca^{2+}$ (f) concentrations for CON (closed circles and solid line) and EXP (open circles and dashed line). The values are means ± S.D.

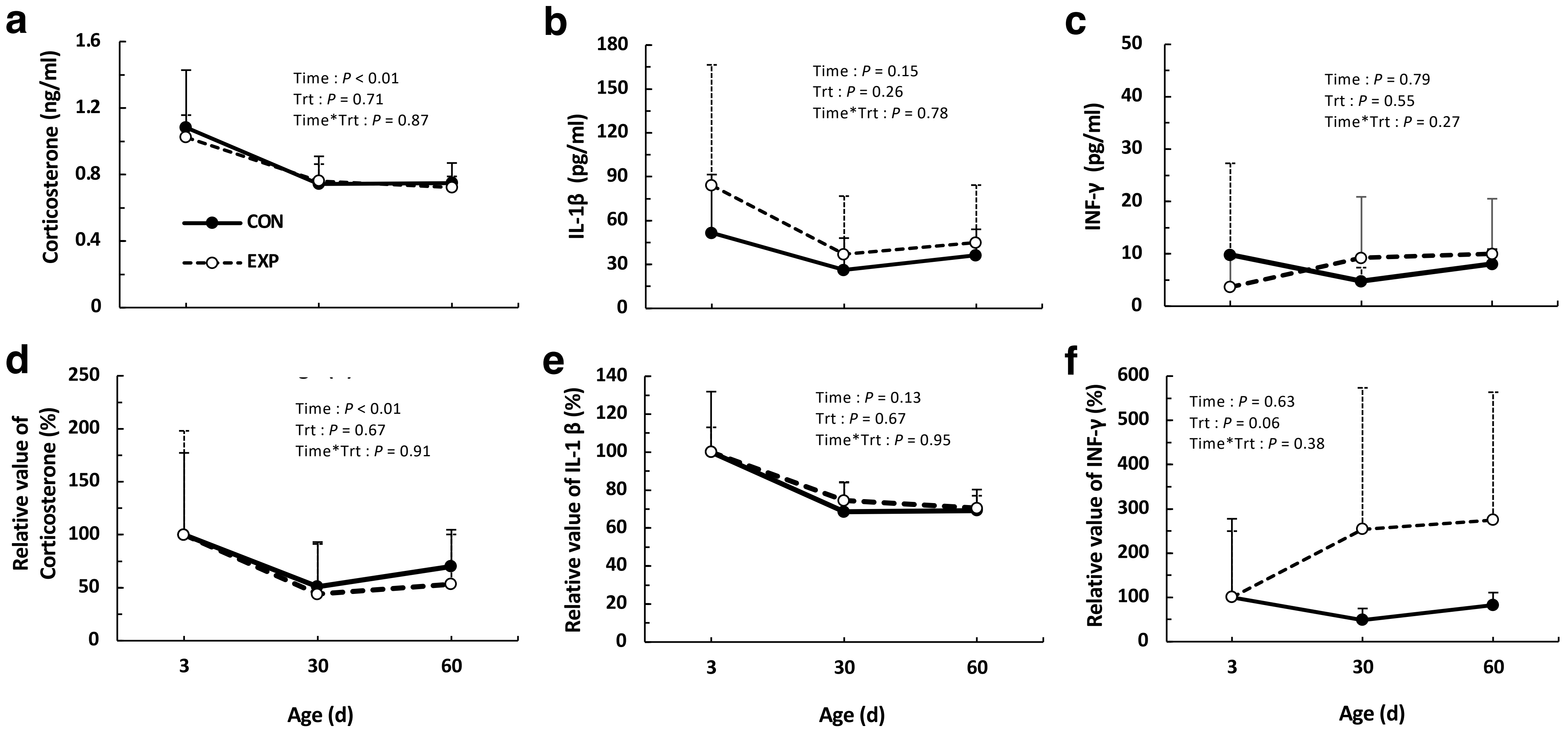

**Figure S3.** Blood components for calves fed milk replacer containing Chlortetracycline (CTC) at 10g/kg (CON) or 0g/kg (EXP). (a-c) Serum concentrations of corticosterone (a), interleukin-1 β (IL- β) (b), interferon-γ (INF-γ) (c). (D-F) Relative values of serum corticosterone (d), IL- β (e) and INF-γ (f) concentrations compared with baseline (3d of age) for CON (closed circles and solid line) and EXP (open circles and dashed line). The values are means ± S.D.

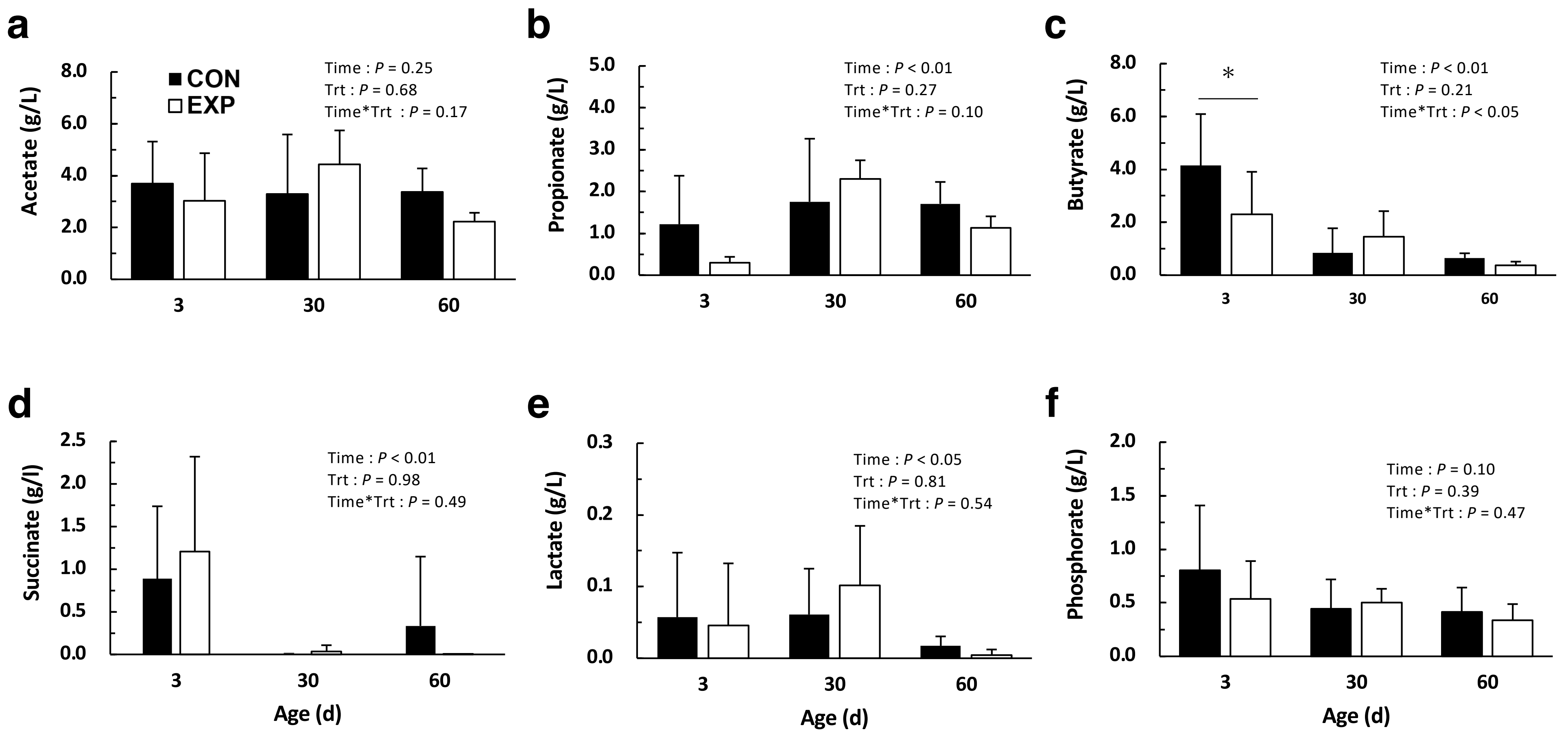

**Figure S4.** Fecal acetate (a), propionate (b), butyrate (c), succinate (d), lactate (e), and phosphate concentrations (f) at 3, 30, and 60 d of age for calves fed milk replacer containing Chlortetracycline (CTC) at 10g/kg (CON) or 0g/kg (EXP). The values are means ± S.D.

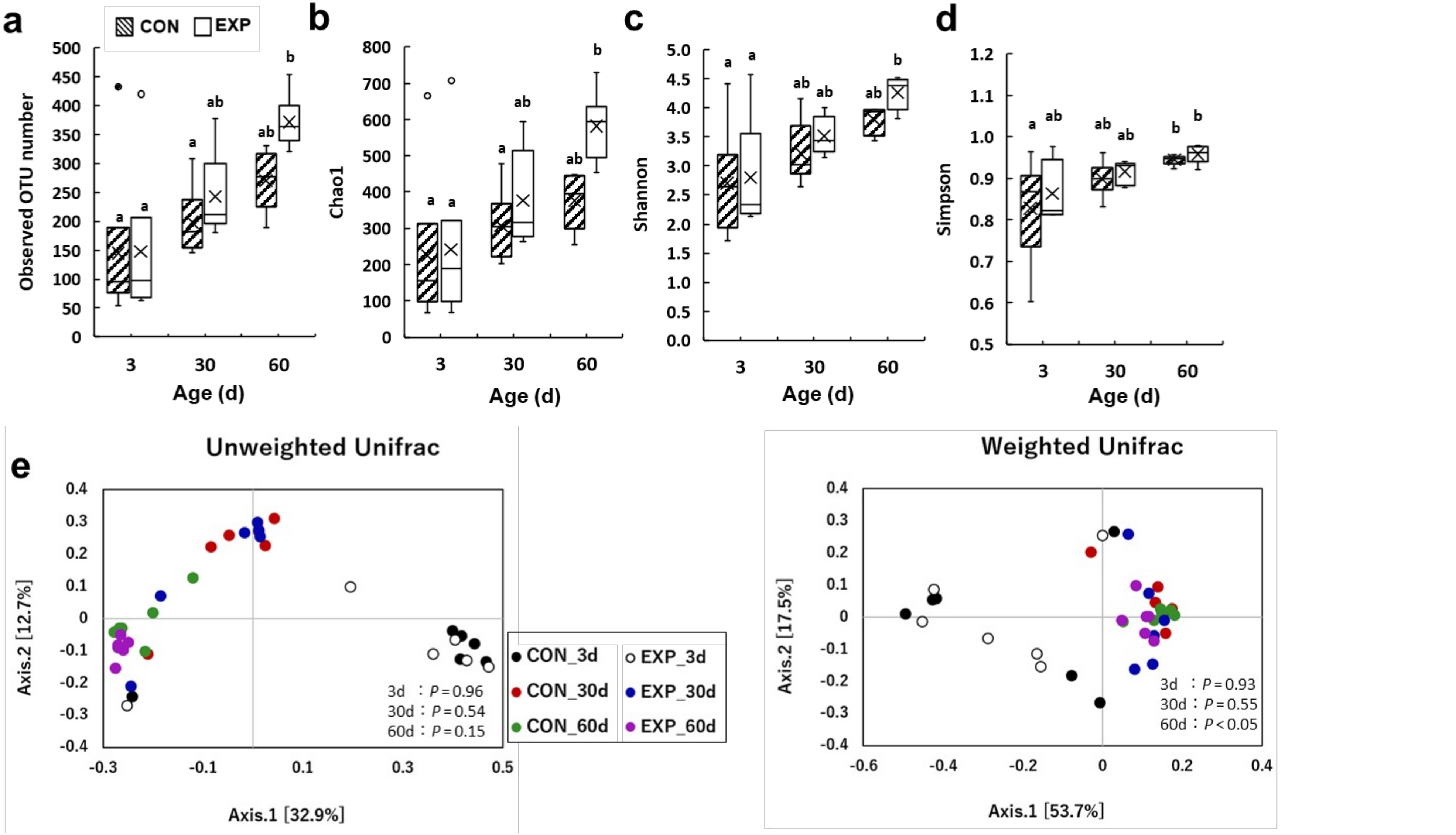

**Figure S5.** Alpha and beta diversity indices for the calves fed milk replacer containing Chlortetracycline (CTC) at 10g/kg (CON) or 0g/kg (EXP). (a) –(d)alpha diversity indices, including observed OTU number (a), Chao1 (b), Shannon (c) and Simpson (d) at 3, 30 and 60 days of age. Beta Diversity indices for the calves fed milk replacer containing CTC at 10g/kg (CON) or 0g/kg (EXP). (e) Indices for β-diversity were estimated using Unifrac analysis with weighted and unweighted principal coordinate analysis (PCoA). Statistic results are shown as follows; Unweighted Unifrac 3d: $R^2$=0.037 P =0.96, 30d: $R^2$=0.083 P = 0.54 and 60d: $R^2$=0.114 P = 0.15, and Weighted Unifrac 3d: $R^2$=0.030 P = 0.93, 30d: $R^2$=0.073 P = 0.55 and 60d: $R^2$= 0.207 $P < 0.05$. Each data was expressed as standard boxplots with medians and averages. Outliers are shown as dots. Different letters above boxplots within a treatment group show significant difference (Tukey's HSD $P < 0.05$). The values are means ± S.D.

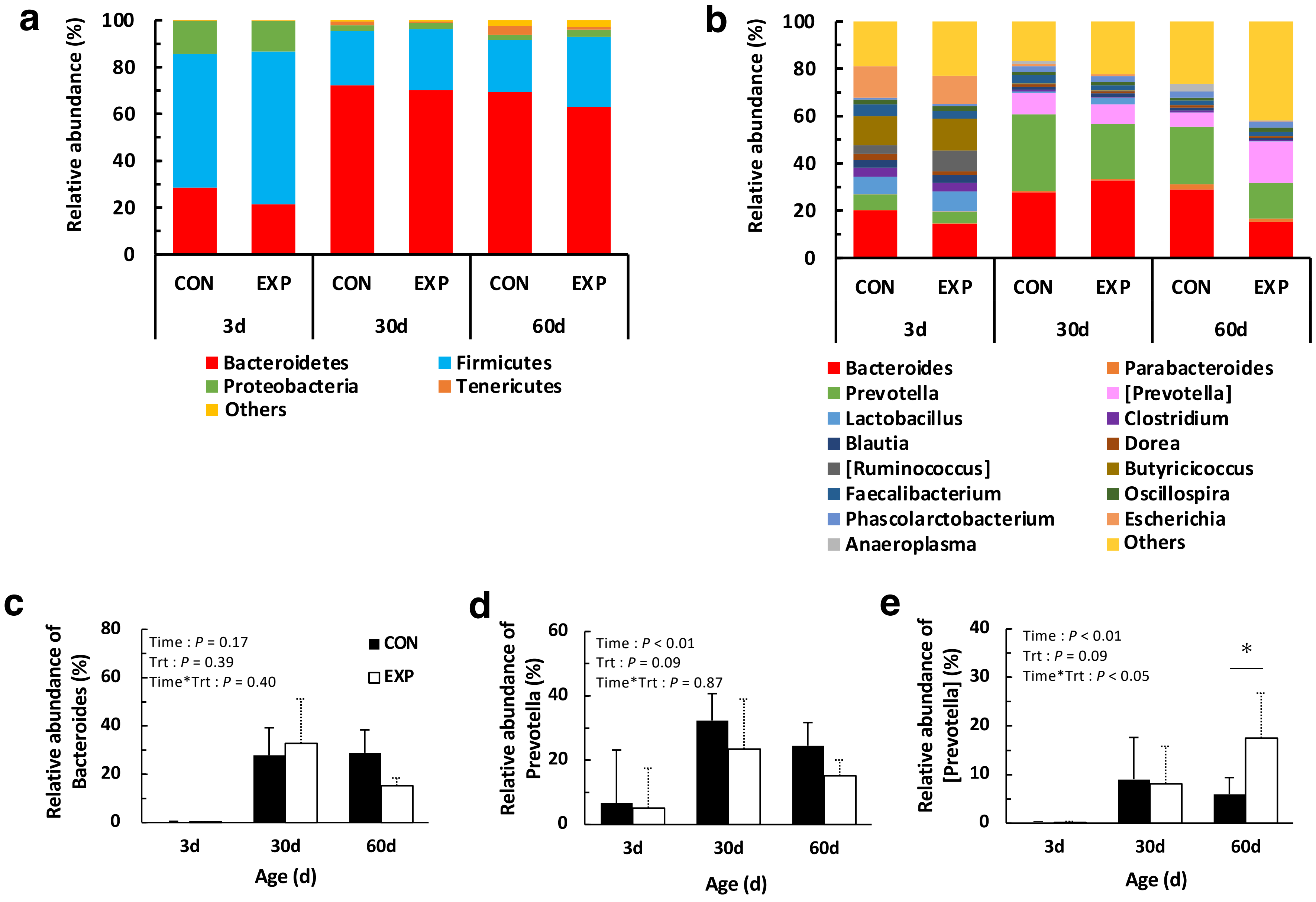

**Figure S6.** (a,b) Relative abundance of fecal microbiota at phylum (a) and genus level (b) for calves fed milk replacer containing Chlortetracycline (CTC) at 10g/kg (CON) or 0g/kg (EXP). (c-e) Relative abundance of genus Bacteroides (c), Prevotella (d), and Prevotella-related bacteria (e) for the CON and EXP.

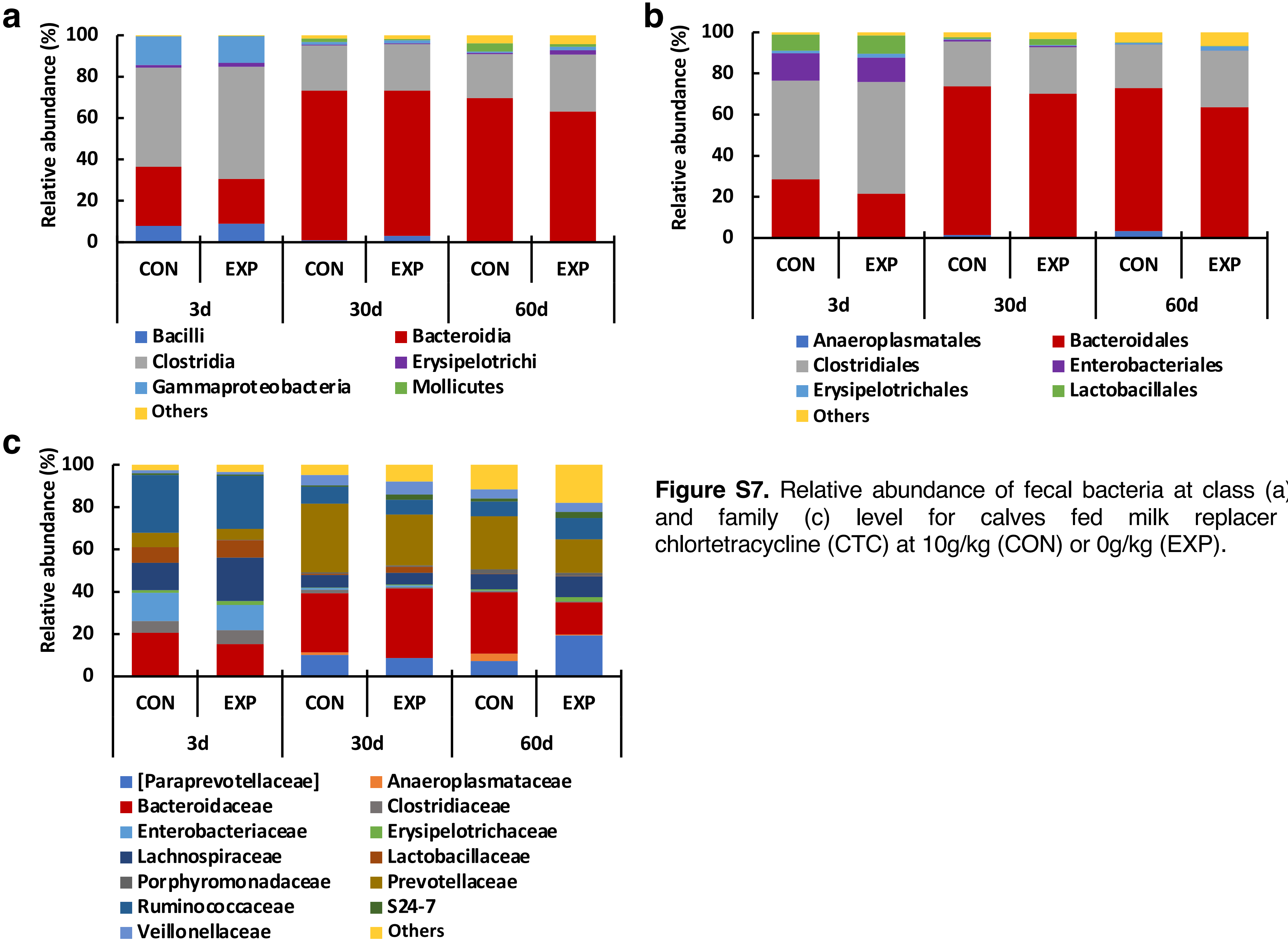

**Figure S7.** Relative abundance of fecal bacteria at class (a), order (b), and family (c) level for calves fed milk replacer containing chlortetracycline (CTC) at 10g/kg (CON) or 0g/kg (EXP).

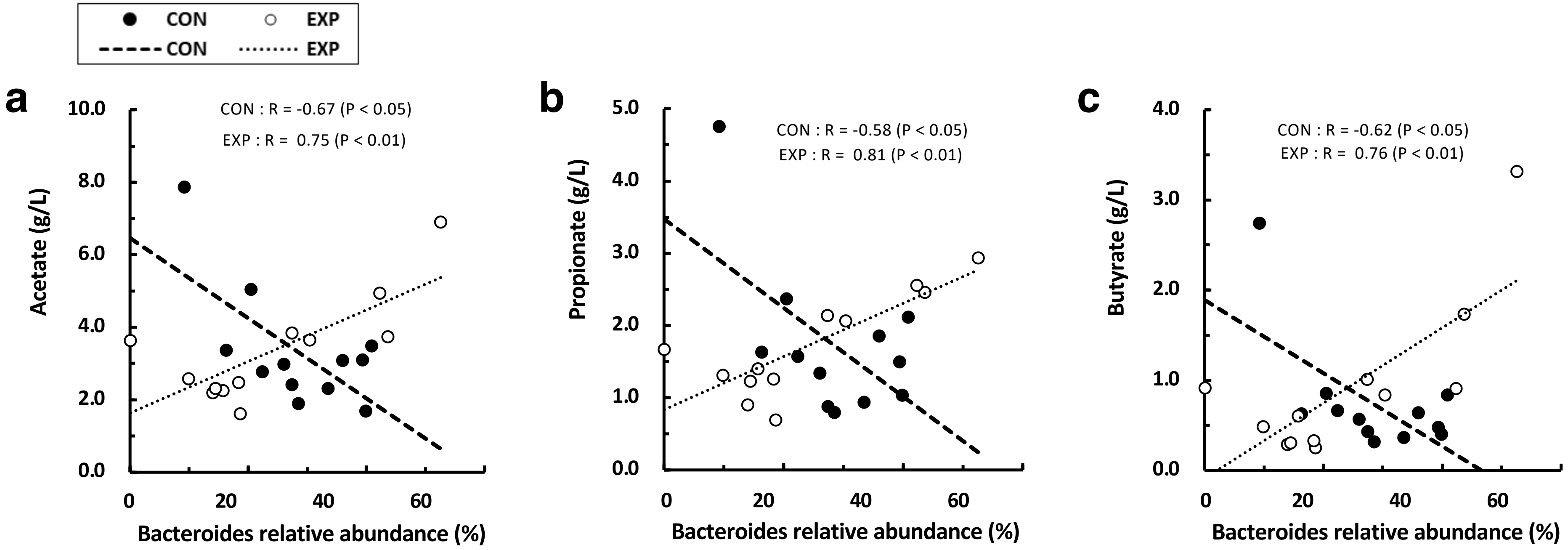

**Figure S8.** Correlation coefficient (r) between fecal abundance of genus Bacteroides and fecal concentrations of acetate (a), propionate (b) and butyrate (c).

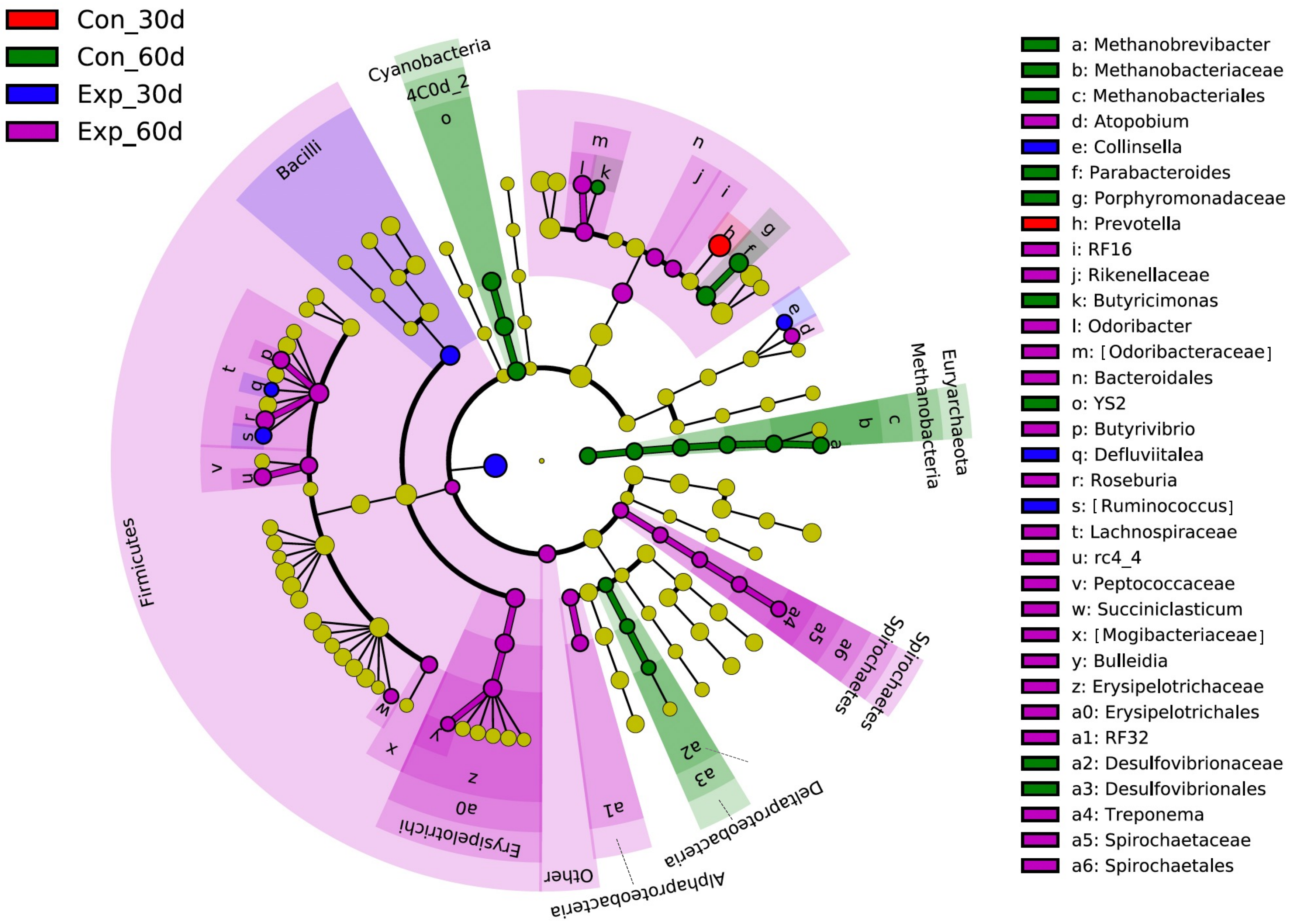

**Figure S9**

LEfSe cladogram visualized based on the significant changes of the bacterial population calculated by the discriminant analysis ($p<0.05$; > 3 fold change).

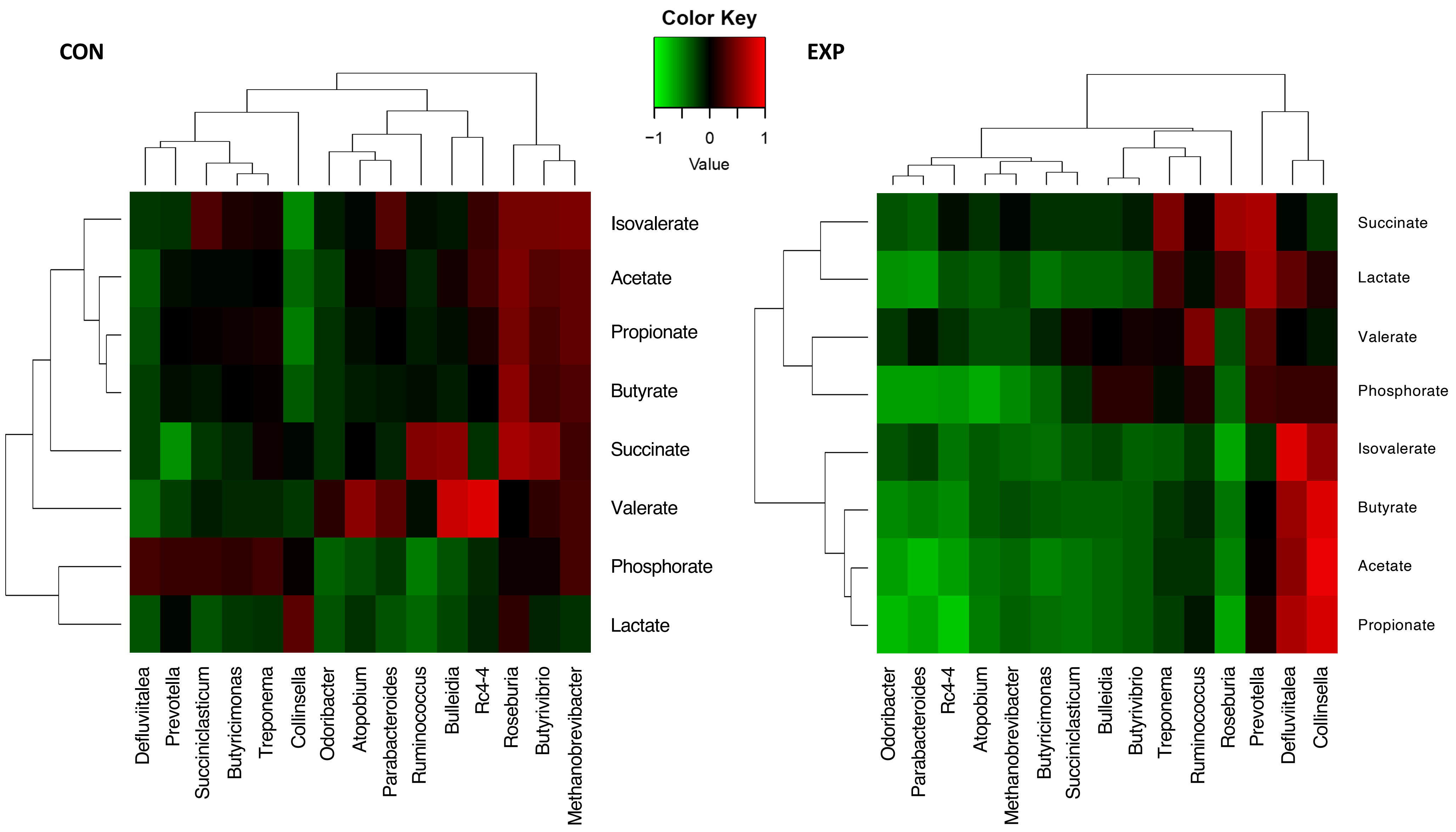

**Figure S10** Heatmaps of correlations between the levels of short-chain fatty acids, lactate, succinate and phosphate and fecal bacterial abundance at the genus level selected by LDA. CON: the group treated with antibiotics; EXP: the group treated without antibiotics.

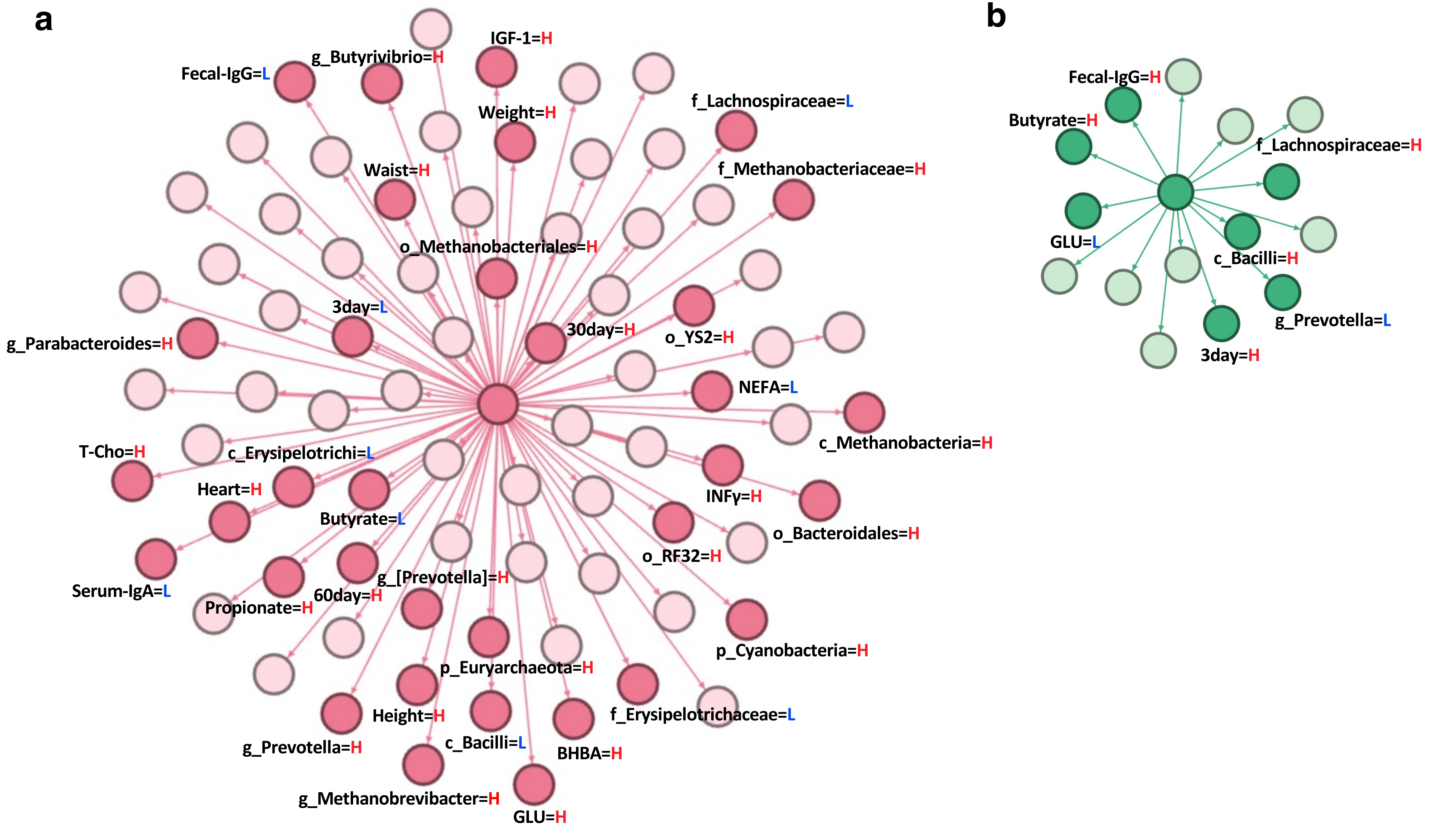

**Figure S11** Association networks of factors sorted (a) with antibiotic treatment and (b) without antibiotic treatment. Positive and negative associated relationships are divided into high (_H) (red color) or low (_L) (blue color) levels based on the mediation values of the whole dataset of targeted components. g_: genus; f_: family; o_: order; c_: class; p_: phylum; Heart: Heart girth_Physique; Height: Withers height_Physique; Waist: Waist circumference_Physique; Weight: Body weight_Physique; GLU: serum glucose; BHBA: β-hydroxybutyric acid; NEFAs: serum nonesterified free fatty acids; T-Cho: serum total cholesterol; IGF-1: Insulin-like growth factor 1; INFγ: interferon γ; IgA: immunoglobulin A; IgG: immunogloblin G.

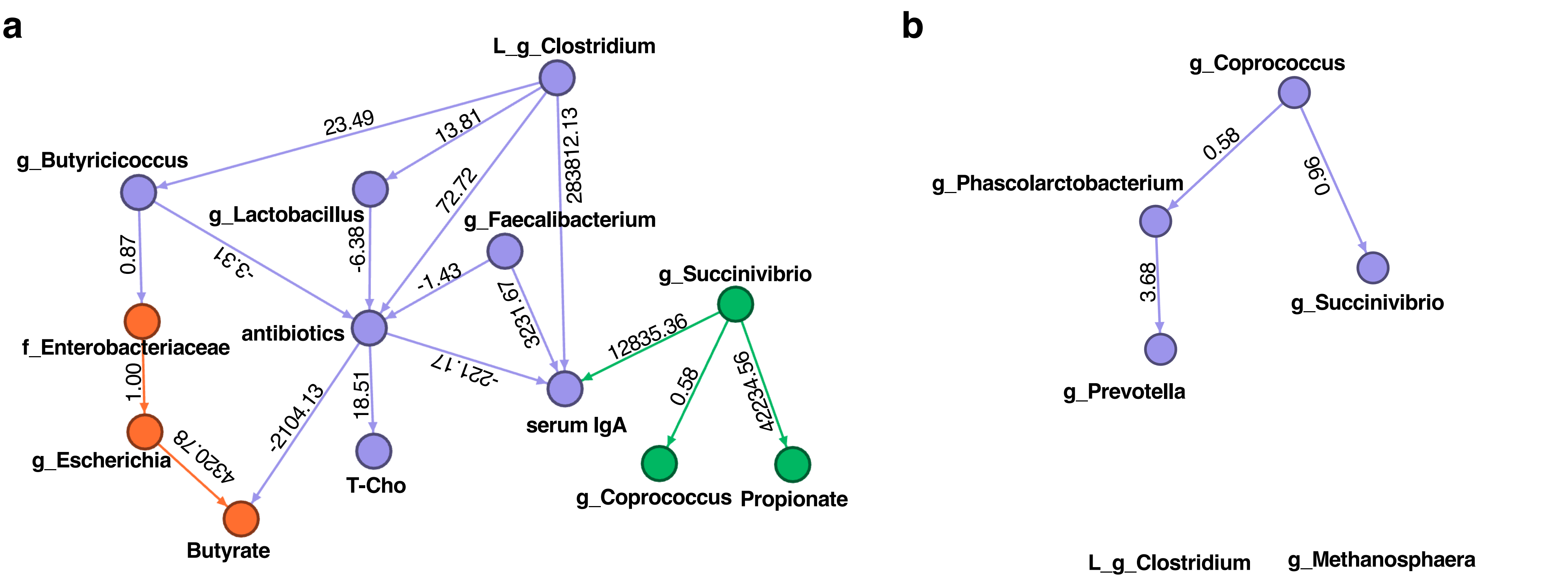

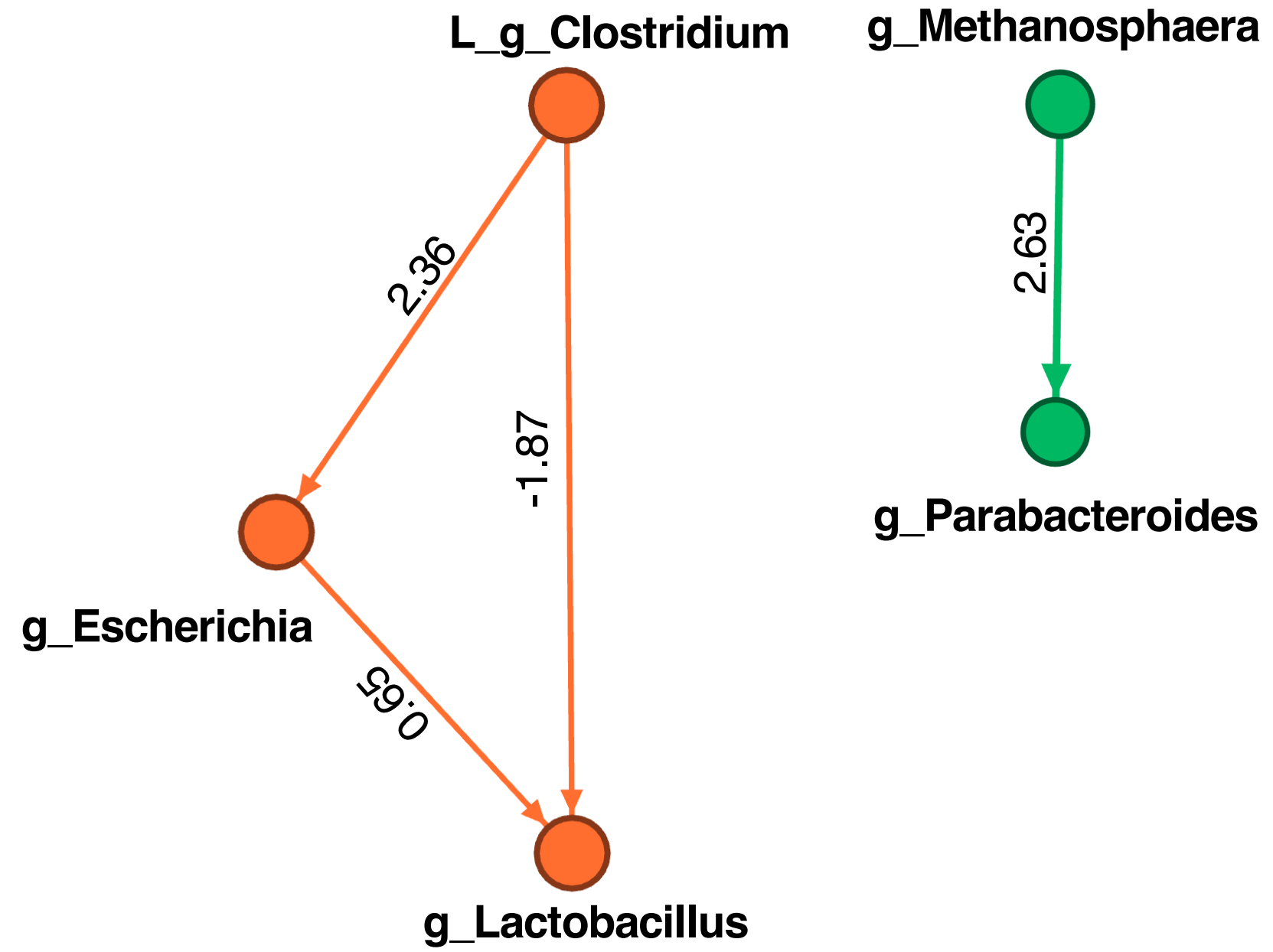

**Figure. S12** DirectLiNGAM-caluculated results for antibiotics-positive components selected by association analysis are shown. All data (family and genus) in antibiotics group (CON) (a) and in non-antibiotics group (EXP) (b) of Days 3, 30, and 60 (> lift value 1.3 in the association analyses) were used. The paths were visualized by the Gephi. The arrow shows a trend of the causal relationship. The number shows the value of the causal contribution (cost) calculated by DirectLiNGAM. The plus and minus value shows positive and negative causal contribution, respectively. The abbreviation show as follows: g_, genus; f_: family; L_: family Lachnospiraceae; bast, bast_Physique; waist, waist_Physique ; T-Cho, total choresterol; IgA: immunoglobulin A.

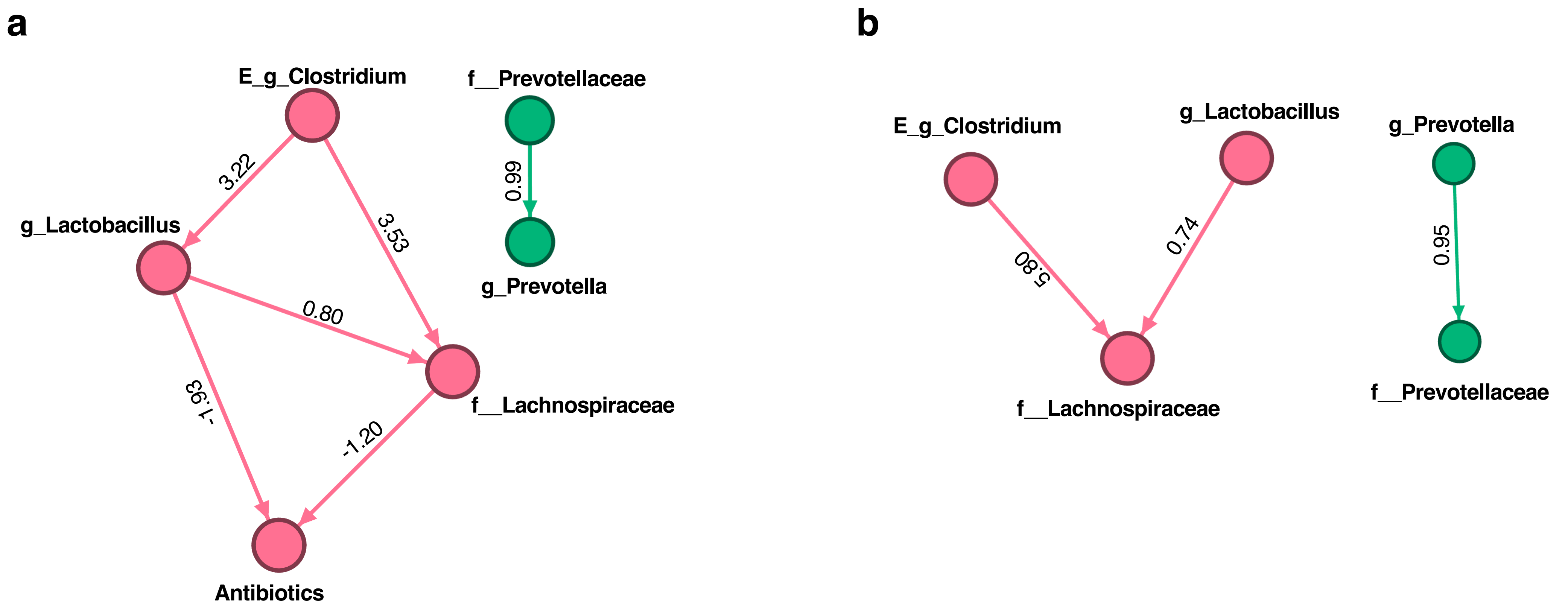

**Figure. S13** DirectLiNGAM-caluculated results for antibiotics-negative components selected by association analysis are shown. All data (family and genus) in antibiotics group (CON) (a) and in non-antibiotics group (EXP) (b) of Days 3, 30, and 60 (> lift value 1.3 in the association analyses) were used. The paths based on all the data were visualized by the Gephi, respectively. The arrow shows a trend of the causal relationship. The number shows the value of the causal contribution (cost) calculated by DirectLiNGAM. The plus and minus cost shows positive and negative causal contribution, respectively. The abbreviation show as follows: g_, genus; f_: family; E_: family Erysipelotrichaceae.

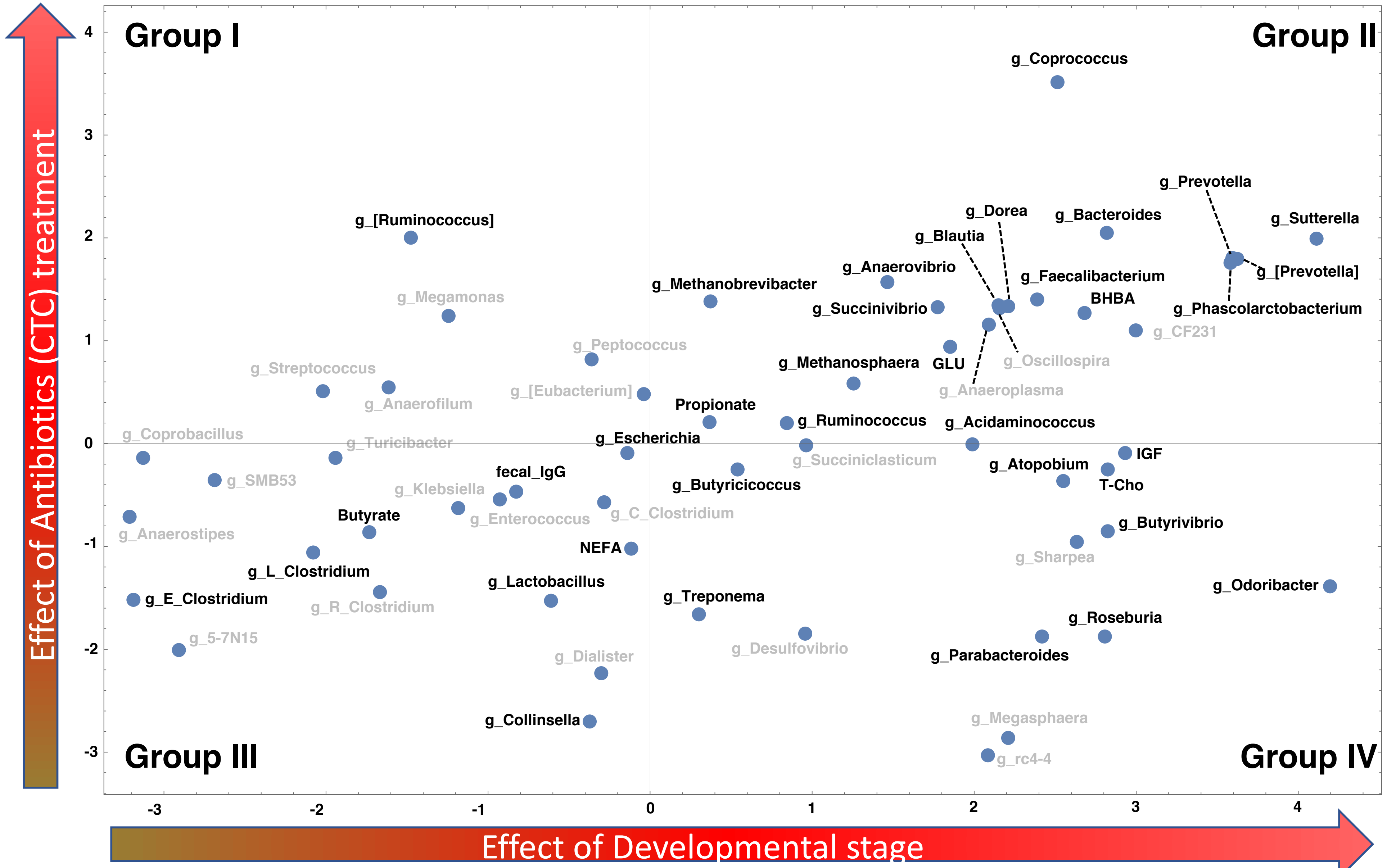

**Figure S14** Response to Environmental ε. Dependencies to developmental stage ($g_i^d$) (X axis) and responsiveness to antibiotics treatment ($g_i^a$) (Y axis) were plotted. Four categories were shown as Group I-IV. The bacteria categorized within Group I have low level of population at 30-60d but increased by antibiotics treatment. The ones within Group II have high level of population at 30-60d and increased by antibiotics treatment. The ones within Group III have low level of population at 30-60d but decreased by antibiotics treatment. The ones within Group IV have high level of population at 30-60d, and/or independent population upon the stage, but the populations decreased by antibiotics treatment. See Fig. 2 and Figures S11-S13 for abbreviation.

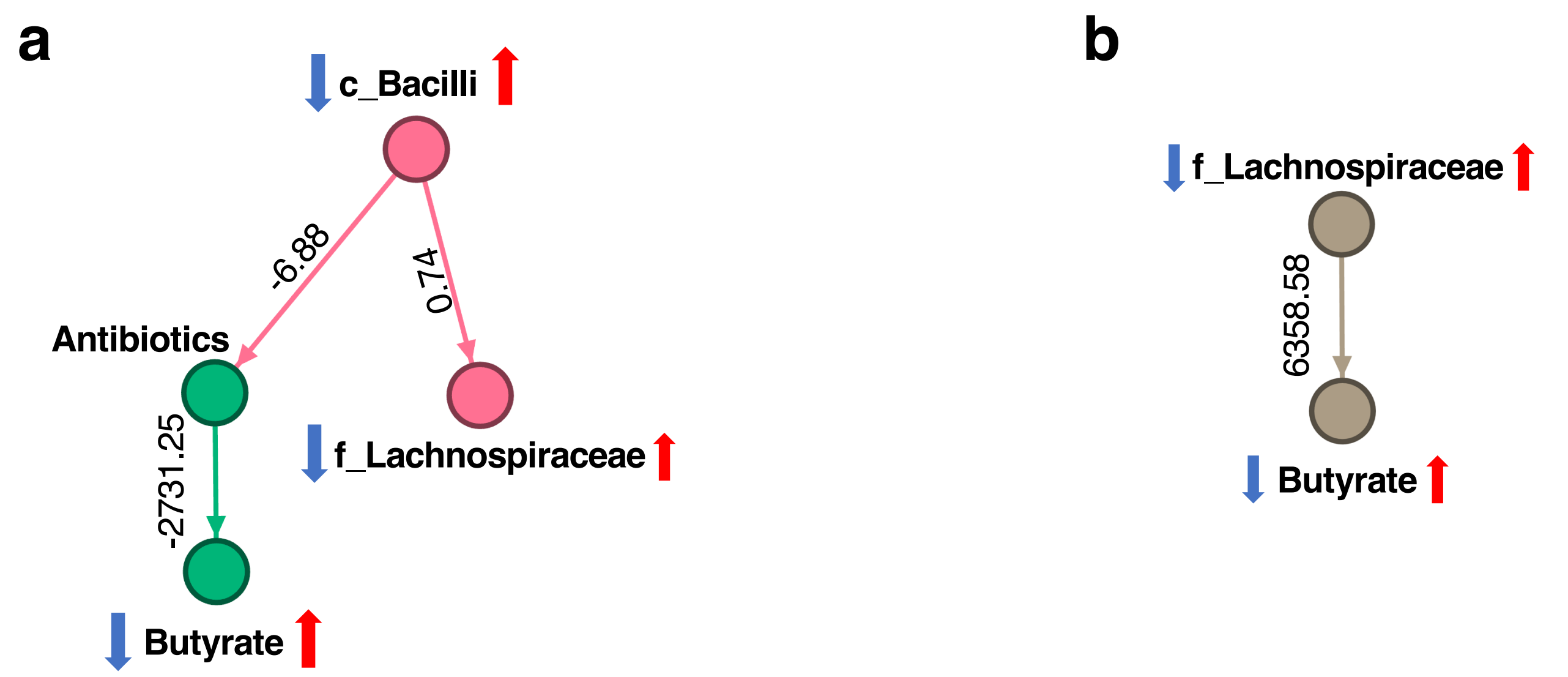

**Figure. S15** DirectLiNGAM-caluculated results of the components associated with class Bacilli (> lift value 1.3 in the association analyses) in group without antibiotics are shown. All data (family and genus) in antibiotics group (CON) (a) and in non-antibiotics group (EXP) (b) of Days 3, 30, and 60 (> lift value 1.3 in the association analyses) were used. The arrows of the right and left side shows the trends of the targeted components in the case with and without antibioitcs, respectively. The upper (red) and lower (blue) arrows show the increased/decreased trend as shown in the association analysis (> lift value 1.3), respectively. The abbreviation show as follows: f_, family, c_ class.

# Supplementary Methods

## Antibiotic-dependent instability of homeostatic plasticity for growth and environmental load

Shunnosuke Okada[♯], *Yudai Inabu*[1], *Hirokuni Miyamoto*[*2,3,4,5], *Kenta Suzuki*[6], *Tamotsu Kato*[3], *Atsushi Kurotani*[7,8], *Yutaka Taguchi*[1], *Ryoichi Fujino*[1], *Yuji Shiotsuka*[1], *Tetsuji Etoh*[1], *Naoko Tsuji*[5], *Makiko Matsuura*[2,5], *Arisa Tsuboi*[2,4,5,8], *Hiroaki Kodama*[2], *Hiroshi Masuya*[6], *Jun Kikuchi*[7], *Hiroshi Ohno*[3]*, *Hideyuki Takahashi*[1]*

*Affiliations:*

[1]*Kuju Agricultural Research Center, Graduate School of Agriculture, Kyushu University, Oita, Japan, 878-0201*

[2]*Graduate School of Horticulture, Chiba University, Chiba, Japan, 263-8522*

[3]*RIKEN Integrated Medical Science Center, Yokohama, Kanagawa, Japan, 230-0045*

[4]*Japan Eco-science (Nikkan Kagaku) Co., Ltd., Chiba, Japan, 260-0034*

[5]*Sermas, Co., Ltd., Chiba, Japan, 271-8501*

[6]*RIKEN, BioResource Research Center, Tsukuba, Ibaraki, Japan, 305-0074*

[7]*RIKEN Center for Sustainable Resource Science, Yokohama, Kanagawa, Japan, 230-0045*

[8]*Research Center for Agricultural Information Technology, National Agriculture and Food Research Organization, Tsukuba, Ibaraki, Japan, 305-0856*

[9]*Feed-Livestock and Guidance Department, Dairy Technology Research Institute, The National Federation of Dairy Co-operative Associations (ZEN-RAKU-REN), Fukushima, Japan*

## Methods

### *Analyses of serum and fecal IgA, IgG and IFN-γ concentrations*

One hundred milligrams of feces was soaked in 1 ml PBS at room temperature for 1 h and centrifuged for 10 min at 10000 × g. After centrifugation, the supernatants were collected for fecal extraction. Total immunoglobulin A (IgA) and IgG concentrations in serum and fecal samples were determined by employing quantitative sandwich ELISA. Ninety-six microtiter plates (C96 Maxisorpcert, Nunc-Immuno Plate, Thermo Fisher Scientific) were directly coated overnight at 4 °C with 0.33 µg/well capture IgA antibody (A10-131A, Bethyl Laboratories, Inc., Montgomery, TX, USA) or 0.2 µg/well capture IgG antibody (A10-118A, Bethyl Laboratories Inc.). After incubation, wells were washed with Tris-buffered saline-Tween-20 (TBST) and then incubated with 100 µL of serum (IgA = diluted 1:10000, IgG = diluted 1:100000 in PBS) or feces (IgA = diluted 1:10, IgG = diluted 1:1 in PBS) for 90 min at room temperature. After five TBST washes, wells were incubated with horseradish peroxidase-conjugated sheep anti-bovine IgA antibody (diluted 1:30000, A10-131P, Bethyl Laboratories, Inc.) or horseradish peroxidase-conjugated sheep anti-bovine IgG antibody (diluted 1:100000, A10-118P, Bethyl Laboratories, Inc.) for 2 h at room temperature. Freshly prepared substrate was added, and the OD was measured at 450 nm using a 3′,3′,5,5′-tetramethylbenzidine microwell peroxidase substrate system (KPL, Gaithersburg, MD, USA) with a spectrophotometer (Multiskan GO, Thermo Fisher Scientific). All samples were analyzed in duplicate, and mean values were calculated. Final IgG and IgA concentrations (mg/mL) in serum and feces were calculated using the standard curve generated for each individual assay taking dilutions into account. Serum and fecal concentrations of interferon γ (IFN-γ) were measured using an ELISA kit (3119-1H-20, Mabtech, Nacka Strand, Sweden) according to the manufacturer's instructions.

### *Association analysis*

Association analysis, an elementary method of unsupervised learning used for market research and ecological analysis, is applied to achieve an understanding beyond the logic of numbers using relative numbers [1-4] as a suitable approach when the classification of data is different for each categorized layer. Here, all growth-related data, metabolomic data, and bacterial populations were analyzed and classified into associated components by subjecting them to conditions in which it is difficult to make horizontal comparisons.

In brief, effects from causes were classified as x and y, respectively. The probability (P) is defined as follows: support, $(x \Rightarrow y) = P(x \cap y)$; confidence, $(x \Rightarrow y) = P(x \cap y)/P(x)$; lift, $(x \Rightarrow y) = P(x \cap y)/P(x)P(y)$. A value of > 1 represents a positive association (if the value indicates independence), a value of < 1 represents a negative association.

Here, association rules were determined by using criterion values of support, confidence, and lift ("support = 0.2, confidence = 0.4, maxlen = 2" and "lift > 1.3"). The data combined with all information, such as body conditional and physiological data, fecal metabolites, and bacterial taxa obtained with or without antibiotic treatment (set as 1 or 0), were used in the analysis. To avoid the differences dependent upon the layers, all the data for the analysis were calculated for binarization based on the median value (M) within the data and sorted as 0 (< M) and 1 (> M). The packages "arules" and "aruleViz" in R software (https://cran.r-project.org) were applied. The association systemic network was rendered by Force Atlas with Noverlap in Gephi 0.9.2 (http://gephi.org).

## Supplementary Text

### *Analyses based on Data S1*

All raw data are summarized in the sheet name "whole raw data", "Fig. S1a", and "Fig. S1b" of DataS1. Based on these data, the optimized data for figure creation are shown as the sheet name as follows: Figs. 1d and S9, Figs.2ab and S15 components, Fig S1a, Fig. S11 Raw data, Fig.S12a, Fig.S12b, Fig.S13a, Fig.S13b, Fig.S14a, and Fig.S14b. The lists in sheet names "Fig.S11a" and "Fig.S11b" of Data S1 show the MBA-calculated raw data in Figs. S11a and S11b, respectively. The list in sheet names "Fig.2a", "Fig.2c", and "S15" shows the raw data of energy landscape analysis, respectively

## References

[1] Shiokawa, Y., Misawa, T., Date, Y. & Kikuchi, J. Application of Market Basket Analysis for the Visualization of Transaction Data Based on Human Lifestyle and Spectroscopic Measurements. *Analytical Chemistry* **88**, 2714-2719 (2016).

[2] Shiokawa, Y., Date, Y. & Kikuchi, J. Application of kernel principal component analysis and computational machine learning to exploration of metabolites strongly associated with diet. *Sci Rep* **8**, 3426 (2018).

[3] Wei, F., Sakata, K., Asakura, T., Date, Y. & Kikuchi, J. Systemic Homeostasis in Metabolome, Ionome, and Microbiome of Wild Yellowfin Goby in Estuarine Ecosystem. *Sci Rep* **8**, 3478 (2018).

[4] Miyamoto, H. et al. A potential network structure of symbiotic bacteria involved in carbon and nitrogen metabolism of wood-utilizing insect larvae. *Sci Total Environ* **836**, 155520 (2022).